\documentclass[preprint,pra,showpacs,eqsecnum]{revtex4}

\usepackage{bm}
\usepackage{amsmath}
\usepackage{graphicx}

\newcommand\vect[1]{\mathbf{#1}}

\begin{document}

\title{Nanostratification of optical excitation in 
self-interacting 1D~arrays}

\author{A.~E.~Kaplan}
\email{alexander.kaplan@jhu.edu}
\author{S.~N.~Volkov}
\affiliation{Dept.\ of Electrical and Computer Engineering, 
Johns Hopkins University, Baltimore, MD 21218}

\date{June 03, 2009}

\begin{abstract}
The major assumption of the Lorentz--Lorenz theory about uniformity of 
local fields and atomic polarization in dense material does not hold 
in finite groups of atoms, as we reported earlier 
[A.~E.\ Kaplan and S.~N.\ Volkov, 
Phys.\ Rev.\ Lett.\ \textbf{101}, 133902 (2008)].
The uniformity is broken at sub-wavelength scale, where the system may 
exhibit strong stratification of local field and dipole polarization, 
with the strata period being much shorter than the incident wavelength.
In this paper, we further develop and advance that theory for the most 
fundamental case of one-dimensional arrays, and study nanoscale 
excitation of so called ``locsitons'' and their standing waves (strata) 
that result in size-related resonances and related large field 
enhancement in finite arrays of atoms.
The locsitons may have a whole spectrum of spatial frequencies, 
ranging from long waves, to an extent reminiscent of ferromagnetic 
domains, -- to super-short waves, with neighboring atoms alternating 
their polarizations, which are reminiscent of antiferromagnetic spin 
patterns.
Of great interest is the new kind of ``hybrid'' modes of excitation, 
greatly departing from any magnetic analogies.
We also study differences between Ising-like near-neighbor 
approximation and the case where each atom interacts with all other 
atoms in the array.
We find an infinite number of ``exponential eigenmodes'' in the 
lossless system in the latter case.
At certain ``magic'' numbers of atoms in the array, the system may 
exhibit self-induced (but linear in the field) cancellation of resonant 
local-field suppression.
We also studied nonlinear modes of locsitons and found optical 
bistability and hysteresis in an infinite array for the simplest modes.

This paper is now published, with very minor changes, in 
Phys.\ Rev.~A \textbf{79}, 053834 (2009).
\end{abstract}



\maketitle
\section{Introduction}
\label{sec1}

This paper is a theoretical extension of 
our recent Letter \cite{cit1} on nanoscale
stratification of local field and atomic dipole
excitation in low-dimensional lattices
driven by a laser at the frequency 
of the resonant atomic transition.
We focus here on the most fundamental 
case of a one-dimensional (1D) array of resonant atoms,
and construct a detailed theory
of both linear and nonlinear interactions
in the system, resulting in such phenomena as 
sub-wavelength spatial modulation 
(stratification) of polarization and local field,  
long-wave and short-wave stratification,
size-related resonances and large field
enhancement, ``magic'' numbers,
ferromagnetic- and antiferromagneticlike
atomic polarization, optical bistability and hysteresises, etc.
In addition to the results \cite{cit1} and their derivations,
we also present new results 
(i) on the size-related resonances using
a many-body approximation involving
interactions of each atom with all other
atoms in the array, beyond the Ising-like approximation 
whereby atoms only interact with their nearest neighbors,
(ii) on traveling locsiton waves and their dissipation,
as well as an estimate of the maximum size
of the 1D array to support most of the effects discussed here,
(iii) on hysteresises and optical bistability in 
arbitrarily long arrays, and
(iv) general mathematical consideration of dispersion
relation in the self-interacting arrays
for dipole approximation and beyond it.

It is well known that optical properties of sufficiently dense 
materials are substantially affected by the near-field interactions
between neighboring particles at the 
frequency of the incident field, in particular, quasi-static (non-radiative)
dipole interactions. 
The best known manifestation of this fact is the
\emph{local vs incident field} phenomenon and 
related Lorentz--Lorenz or Clausius--Mossotti relations \cite{cit2}
for dielectric constant as a nonlinear function of the number density.
The microscopic (or \emph{local})
field (LF), $\vect{E}_\mathrm{L}$, acting upon atoms or molecules
becomes then different from both the applied and
average macroscopic fields because of inter-particle interaction.
In the most basic case, that relation between
$\vect{E}_\mathrm{L}$ and the average field $\vect{E}$ is 
$\vect{E}_\mathrm{L} = (\epsilon +  2) \vect{E} / 3$,
where $\epsilon$ is the dielectric constant of the material 
at the laser frequency $\omega$.
It is worth noting that we are not interested here in the relation
between $\epsilon$ and the number density of the material, since in
the case of small 1D or 2D arrays the issue is moot.
For the same reason, it also makes sense 
to deal directly with the incident field, $\vect{E}_\mathrm{in}$,
instead of the averaged field.
In most theories of those interactions a
traditional standard (and at that often implicit) assumption,
reflected also in the above formula, 
is that the local field and polarization
are uniform in the near neighborhood of each particle
at least at the distances shorter than the wavelength
of light, $\lambda$.
This essentially amounts to 
the so called ``mean-field approximation''.

It was shown by us \cite{cit1}, however, that if the 
\emph{local uniformity} is not presumed, then, under certain 
conditions on the particle density and their dipole strengths,
the system of interacting particles is bound to exhibit 
periodic spatial variations of polarization and local-field 
amplitude.
These variations result in sub-wavelength 
strata with a nanoscale period  much shorter than $\lambda$.
Under certain conditions, the system may exhibit
an ultimate non-uniformity, whereby
each pair of neighboring atoms in 1D arrays has their dipoles
counter-oscillating with respect to each other,
i.\,e.\ their excitations and thus local fields
have opposite signs.

To a certain extent this is reminiscent of the situation
in magnetic materials
with \emph{ferromagnetic vs antiferromagnetic} effects.
Indeed, the mean-field approximation which is
at the root of the Curie--Weiss theory of 
\emph{ferromagnetism} \cite{cit3}, 
is based on the assumption
of uniform polarization of all the neighboring
magnetic dipoles even without external
magnetic field, when taking
into account their interaction with each other.
Contrary to that, the Ising theory \cite{cit3},
which does not make the assumption of uniformity, showed that 
even in the near-neighbor approximation for that interaction,
the excitation may result in counter-polarization
of neighboring atoms in 1D arrays, and thus in a
new phenomenon of \emph{antiferromagnetism}.
Note here that in these effects
there is no notion of ``local'' \emph{vs} ``external''
field phenomenon: in a ``pure'' case of either
of the magnetic effects, no external field is applied;
the effects here are the result of self-organization
of permanent, ``hard'', atomic dipoles with 
pre-existing \emph{dc} dipole fields, without
any ``help from outside''.

In this lies a profound difference between 
\emph{dc} magnetic material phenomena (ferro- and antiferromagnetism)
on the one hand -- and the effects considered by us here and in \cite{cit1}
on the other hand, all of which are based on the \emph{optical} 
(or, in general, any other quantum or classical resonance)
excitation of atoms (or other small particles, e.\,g.\ quantum dots,
clusters, small-particle plasmons, etc.).
While magnetic dipoles in ferromagnetics
are nonzero even in the absence
of an external field (we may call them ``hard'' dipoles), 
the oscillating dipoles (in the linear case)
can be induced only by the driving field 
at the near-resonant frequency, 
so they can be called ``soft'' dipoles; 
without such a driving their polarization vanishes.
Because in dense material
the atoms actually are acted upon by \emph{local}
field, the response of
each one of them may differ from the others
by phase and amplitude (or even direction),
with some of the dipoles fully suppressed
while others fully excited.
Thus the effects considered here are induced 
by the interplay of external and local fields, which
put the entire phenomenon squarely into
the domain of relations between the local and incident fields.
Because of that, since the phenomenon
depends strongly on the characteristics of the 
incident field (polarization, frequency, and, 
in the nonlinear case, intensity),
the spatial modulation of the dipole
excitation and local field can vary substantially.
This results in a wealth of different patterns,
some of them reminiscent of the ferromagnetic,
other of antiferromagnetic, but
the most of them forming all kinds of hybrid patterns.
The complete cross-over from ferromagneticlike to 
antiferromagneticlike state of the system with 
all the intermediate states can be attained then 
by simply tuning laser frequency.

Another significant difference here
is that the system size is small.
Provided there is sufficiently strong interparticle interaction,
the new phenomenon can occur in the vicinity of boundaries,
lattice defects, impurities or in sufficiently small group of atoms;
recent advances in technology allow
fabrication of nanoscale structures with small numbers of atoms.
Thus, our theory emphasizes phenomena in relatively small ordered arrays of
interacting atoms, in contrast to, e.\,g., microscopic
models of ferromagnetism that mostly focus on averaged,
``thermodynamic'' perspective on sufficiently large systems.
This  brings forward a new set of nanoscale phenomena.
Harking back to ferromagnetic systems, this new emphasis
may reveal similar phenomena for nanoscale magnetic
systems, which could be an exciting topic for a separate study.

Our choice of 1D and 2D dielectric systems
based on the arrays or lattices
of atoms, quantum dots, clusters, molecules,
etc., allows to control anisotropy of near-field interaction.
It also eliminates the issues of electromagnetic (EM) propagation being
modified by the effects as the EM wave propagates through the structure
(especially if it propagates normally to the lattice).

If \emph{local uniformity} is broken by any perturbation, the system
may exhibit near-periodic spatial sub-$\lambda$
patterns (strata) of polarization.
In general, two major modes of the strata 
transpire: \emph{short-wave} (SW), with the period up 
to four interatomic spacings, $l_a$, 
and \emph{long-wave} (LW) strata.
The strata are standing waves of elementary
LF excitations (called \emph{locsitons} in \cite{cit1})
having a near-field, electrostatic, nature and low group velocity.

In the first approximation,
the phenomenon is linear in the driving field,
and the locsitons may be excited within a 
spectral band much broader than the atomic linewidth.
It can be viewed as a Rabi broadening of an atomic
line by interatomic interactions.
The strata are controlled by laser polarization and
the strength of atom coupling, $Q$, 
\emph{via} atomic density, dipole moments, relaxation, and detuning.
Once $|Q| > Q_\mathrm{cr} = O(1)$,
the LF uniformity can be broken by boundaries, impurities, 
vacancies in the lattice, etc.
A striking manifestation of the effect
is large field resonances due to locsiton
eigenmodes in finite lattices,
and -- at certain, ``magic'' numbers of atoms in the lattice --
almost complete cancellation of 
field suppression at the atomic resonance;
saturation nonlinearity results 
in hysteresises and optical bistability.

The paper is structured as follows.
In Section~\ref{sec2} we derive the main equations
for self-interacting atomic lattices of arbitrary 
dimensions using two-level (nonlinear in general) model
for atomic resonances 
and dipole--dipole interaction between atoms,
while Section~\ref{sec3} is on specific 
equations for linear infinite and finite 1D arrays.
In Section~\ref{sec4} we develop the general theory
of locsitons and derive the dispersion relation.
In Section~\ref{sec5} we study locsiton band formation,
size-related resonances due to
standing waves of locsitons (strata) and local-field enhancement.
In Section~\ref{sec6} we concentrate on detailed theory
of resonances beyond the near-neighbor approximation,
including evanescent solutions (see also Section~\ref{sec4}).
Magic numbers are considered in Section~\ref{sec7}.
In Section~\ref{sec8}, we study the effects of losses on 
locsiton excitation,
depth of penetration, and traveling  locsiton waves.
Section~\ref{sec9} is on nonlinear locsiton modes, in particular,
optical bistability and hysteresis.
Section~\ref{sec10} addresses potential applications
of locsitons and their analogies
in other physical systems.
In Conclusions, Section~\ref{sec11}, we summarize our results.
Appendix~\ref{app} is on general mathematical aspects
of dispersion relations for 1D arrays.

\section{Main equations}
\label{sec2}

Our model is based on the near-field dipole
atomic interactions, with the incident frequency $\omega$ being
nearly resonant to an atomic transition
with a dipole moment $\vect{d}_a$ at the frequency $\omega_0$.
In the linear case, i.\,e.\ when the laser intensity
is significantly lower than that for the quantum
transition saturation (see below),
the result of this model coincides
with that of a classical harmonic oscillator
formed by an electron in a harmonic potential
with the same resonant frequency $\omega_0$
and with the dipole moment
\begin{equation}
|d_a| = \frac{e}{2 \pi} \sqrt{\lambda_\mathrm{C} \lambda / 2},
\label{eq2.1}
\end{equation}
where $\lambda_\mathrm{C} = 2 \pi \hbar / m c$ is the Compton
wavelength of electron.

In a standard LF situation, $\lambda \gg l_a$,
where $\lambda = 2 \pi c / \omega$,
the field of an elementary dipole with the
polarization $\vect{p}$ is dominated in its near vicinity
by a non-radiative, quasi-static 
(and only electric) component,
which is strongly anisotropic in space.
This dominant term in the near-field area, 
$|\vect{r}' - \vect{r}| = r_0 \ll \lambda$,
attenuates as $1 / r_0^3$ (see e.\,g.\ \cite{cit4}, Section~72).
At $|d_a| \ll r_0 \ll \lambda$ the amplitude 
of the field of an oscillating dipole located
at $\vect{r}'$ induces a field
at the point of observation $\vect{r}$
with the amplitude
coinciding with that of an elementary static dipole with 
the same polarization $\vect{p}'$:
\begin{equation}
\vect{E}_\mathrm{dp} (\vect{r}', \vect{r}) 
= \frac{3 \vect{u} (\vect{p}' \cdot \vect{u}) - \vect{p}'}
       {\epsilon |\vect{r}' - \vect{r}|^3},
\label{eq2.2}
\end{equation}
where $\vect{u} = (\vect{r} - \vect{r}') / r_0$ is
a unit vector in the direction of observation,
and $\epsilon$ is a background dielectric constant.

We will model a resonant atomic transition by a basic two-level
atom in a steady-state mode under the action of a 
field, $\vect{E} \exp( -i \omega  t) / 2 + \mathrm{c.c.}$,
with the amplitude $\vect{E}$ and frequency $\omega$,
and assume that $l_a \gg |d_a|/e$, so that
the wave functions of neighboring atoms
do not overlap.
Using a semi-classical
approach standard in LF theory of resonant atoms \cite{cit5,cit6},
we can now find the atomic polarization as:
\begin{equation}
\vect{p} = - \frac{2 |d_a|^2}{\hbar\Gamma}
    \frac{\vect{E} \Delta N}{\delta + i},
\label{eq2.3}
\end{equation}
where $\vect{p}$ is the polarization amplitude
[the full polarization is then
$\vect{p} \exp( -i \omega t) / 2 + \mathrm{c.c.}$];
$\Delta N = N_1 - N_2$ is the population difference,
with $N_1$ and $N_2$ being 
atomic populations at respective
ground and excited levels ($N_1 + N_2 = 1$), \
$\delta = T \Delta \omega = T (\omega - \omega_0)$
is a dimensionless detuning from
the resonant frequency $\omega_0$
of the two-level atom, $T = 2/\Gamma$ is a
\emph{transverse} relaxation time
(the time of polarization relaxation)
and $\Gamma$ is the (homogeneous) linewidth of the linear 
resonance
\footnote{%
In general, the total homogeneous linewidth $\Gamma = 2/T$ is comprised of
two contributions \cite{cit5}: one is the so called natural
radiation linewidth $\Gamma_a$, which is due
to spontaneous radiation at the resonant frequency of the
respective quantum transition of a single atom in vacuum,
and the other one, $\Gamma_\mathrm{nr}$, is due to all other processes,
mostly phase broadening, non-radiative transitions, and 
radiative ones to non-resonant quantum levels.
It is worth noting that when the atoms are very close to each other,
$l_a \ll \lambda$,
due to atomic interactions
the radiation linewidth, $\Gamma_a$,
of each atom can be broadened up if the neighboring dipoles
move in phase (as in long-wave locsitons),
or narrowed down if those dipoles move in counter-phase
(as in short-wave locsitons).
In general case this effect should be taken into consideration too.
However, we assume here that, as is usual for atoms
in a dense-material environment (e.\,g.\ when they
are positioned at the surface of some material or
embedded in it), $\Gamma_\mathrm{nr} \sim \Gamma \gg \Gamma_a$,
and we ignore here the natural linewidth,
$\Gamma_a$ and its broadening.%
}. 
In turn, the steady-state population difference is \cite{cit5}
\begin{equation}
\Delta N = N^\mathrm{eq} 
    \left( 1 + T \tau \frac{|d_a E/\hbar|^2}{1 + \delta^2} \right)^{-1},
\label{eq2.4}
\end{equation}
where $\tau$ is a \emph{longitudinal} relaxation time
(life-time of the excited atom), and
$N^\mathrm{eq}$ is an equilibrium population difference
at the system temperature due to Boltzmann's distribution;
in optics one can usually assume $N^\mathrm{eq} \approx 1$, so that
\begin{eqnarray}
\Delta N &\approx& 
    \left[ 1 +  \frac{|E|^2 / E_\mathrm{sat}^2}{1 + \delta^2} \right]^{-1} 
\equiv f_\mathrm{NL}(|E|^2),
\nonumber
\\
E_\mathrm{sat}^2 &=& \frac{\hbar^2}{|d_a|^2} \cdot \frac{1}{\tau T},
\label{eq2.5}
\end{eqnarray}
where $E_\mathrm{sat}^2$ is the saturation intensity,
and $f_\mathrm{NL}(|E|^2)$ is nonlinearity due to saturation.
Substituting (\ref{eq2.5}) into (\ref{eq2.3}),
one obtains a closed-form expression for the polarization,
\begin{equation}
\vect{p} = -\vect{E} \frac{2 |d_a|^2}{\hbar\Gamma (\delta + i)} 
    f_\mathrm{NL} (|E|^2).
\label{eq2.6}
\end{equation}
For a classical harmonic (linear) oscillator we have:
\begin{equation}
\vect{p} = -\frac{\vect{E} e^2}{\omega_0 \Gamma m (\delta + i)},
\label{eq2.7}
\end{equation}
where $m$ is the mass of electron.

The local field $\vect{E}_\mathrm{L} (\vect{r})$
at each atom is the incident laser field, $\vect{E}_\mathrm{in}$, 
plus the sum of the near-fields, 
$\vect{E}_\mathrm{dp} (\vect{r}', \vect{r})$ (\ref{eq2.2}), induced 
by all the surrounding dipoles at $\vect{r}'$ acted upon by
the respective local fields $\vect{E}_\mathrm{L} (\vect{r}')$, i.\,e.
\begin{eqnarray}
\vect{E}_\mathrm{L} (\vect{r}) &=& \vect{E}_\mathrm{in} (\vect{r}) 
+ \sum_\mathrm{latt}^{\vect{r}' \ne \vect{r}}
    \vect{E}_\mathrm{dp} (\vect{r}', \vect{r}) 
\nonumber
\\
&=& \vect{E}_\mathrm{in} (\vect{r}) + 
    \frac{1}{\epsilon} \sum_\mathrm{latt}^{\vect{r}' \ne \vect{r}}
    \frac{3\vect{u} (\vect{p}' \cdot \vect{u}) - \vect{p}'}
         {|\vect{r}' - \vect{r}|^3},
\label{eq2.8}
\end{eqnarray}
where $\sum_\mathrm{latt}$ denotes summation over 
the entire lattice or array.
To obtain a closed-form master equation, e.\,g.\ for
$\vect{E}_\mathrm{L} (\vect{r})$ alone, we use (\ref{eq2.6}) to write:
\begin{eqnarray} 
&&
\vect{E}_\mathrm{L} (\vect{r}) = \vect{E}_\mathrm{in} (\vect{r})
- \frac{Q}{4} \sum_\mathrm{latt}^{\vect{r}' \ne \vect{r}}
\left| \frac{l_a}{\vect{r}' - \vect{r}} \right|^3
\nonumber
\\
&&\quad{}\times
\left\{ 3\vect{u} [\vect{E}_\mathrm{L}(\vect{r}') \cdot \vect{u}] 
        - \vect{E}_\mathrm{L}(\vect{r}') \right\}
f_\mathrm{NL} [|\vect{E}_\mathrm{L} (\vect{r}')|^2 ],
\label{eq2.9}
\end{eqnarray}
where $\vect{E}_\mathrm{L}(\vect{r})$ are local fields
only at the locations of atoms in
the lattice and not at any other points 
inside or outside it;
$Q = Q_a / (\delta + i)$ is a tuning-dependent
strength of dipole--dipole interaction, and the maximum absolute
strength, $Q_a$, is
\begin{equation}
Q_a = \frac{8 |d_a|^2}{\epsilon\hbar\Gamma l_a^3} 
= \frac{4\alpha}{\pi\epsilon} \frac{\lambda_0 (|d_a|/e)^2}{l_a^3}
    \frac{\omega_0}{\Gamma},
\label{eq2.10}
\end{equation}
where $\alpha = e^2 / \hbar c \approx 1/137$ is
the fine-structure constant,
and nonlinear factor $f_\mathrm{NL}$ is as in (\ref{eq2.5}).
For a classical harmonic oscillator, (\ref{eq2.1}), we have
\begin{equation}
(Q_a)_\mathrm{class} = \frac{4 e^2}{\epsilon m \omega_0 \Gamma l_a^3}
= \frac{1}{\epsilon \pi^2} \frac{r_e \lambda_0^2}{l_a^3} 
    \frac{\omega_0}{\Gamma},
\label{eq2.11}
\end{equation}
where $r_e \equiv e^2 / m_e c^2 \approx 2.8 \times 10^{-6}$~nm
is the classical radius of electron.

Eqs.\ (\ref{eq2.8}), (\ref{eq2.9}) reflect many-body nature 
of the interaction.
A conventional approach to local fields 
within the Lorentz--Lorenz theory is to look for a self-consistent
solution for the fields in this interaction, 
with an assumption, however, that they are
\emph{uniform} (the mean-field theory), i.\,e.\
to set 
$\vect{E}_\mathrm{L} (\vect{r}) = \vect{E}_\mathrm{L} (\vect{r}')$
and also use an encapsulating sphere around the observation point.
These assumptions effectively shut out
any strong spatial variations of the atomic excitations
and local field that may exist at the inter-atomic scale.  
That is where we depart from the Lorentz--Lorenz
theory; \emph{none of those assumptions are used here},
and our approach is to use general expressions (\ref{eq2.8}) or
(\ref{eq2.9}) and seek straightforward solution for them.

We will see below that the major critical condition
for the phenomenon to exist and be observable
at least at other optimal conditions is
that $Q_a$ exceeds some critical value,
$Q_a > O(1)$.
Three parameters are critical in this respect:
atomic dipole moment $d_a$, the spacing between atoms,
$l_a$, and the atomic linewidth, $\Gamma$,
since $Q_a \propto |d_a|^2 / (\Gamma l_a^3)$.
To get an idea of whether the above critical condition 
is realistic, let us look first
at the case of a gas-like collection 
of atoms, with the relaxed requirement on the spacing $l_a$.
Large dipole moments and narrow resonances in,
e.\,g., \emph{alkali} vapors \cite{cit6} or CO$_2$ gas \cite{cit5},
in solids \cite{cit8}, quantum wells and clusters
may greatly enhance the phenomenon
and allow for $l_a$ from sub-nanometer to a few tens of nanometers.
Considering an example with $l_a \sim 100$~\AA,
corresponding to the volume 
density of $\sim 10^{18}$~cm$^{-3}$,
$|d_a|/e \sim 1$~\AA, \ $\lambda_0 \sim 1$~$\mu\mathrm{m}$,
$\epsilon = 1$, and $\Gamma / \omega_0 \sim 10^{-6}$,
all of which are reasonable data, we obtain $Q_a \sim 10^2$,
which provides a margin large enough to 
see all the effects discussed here.
It is also of interest to roughly estimate
what is the upper limit for $Q_a$.
To that end, consider the extreme situation
of $l_a \sim |d_a|/e$
(solid-state- or liquid-like packing of participating atoms),
in which case we have the ceiling for $Q_a$ as
\begin{equation}
Q_\mathrm{ceil} = \frac{8 e^3}{\epsilon\hbar\Gamma |d_a|} 
= \frac{4\alpha}{\pi\epsilon} \frac{\lambda_0}{(|d_a|/e)}
    \frac{\omega_0}{\Gamma}.
\label{eq2.12}
\end{equation}
Even taking into consideration
significant line broadening, $\Gamma$,
$Q_\mathrm{ceil}$ may exceed unity by many order
of magnitude, thus providing
huge margin for the existence
and observation of locsitons and related effects.

\section{1D array of atoms: linear case}
\label{sec3}

\begin{figure}
\includegraphics[width=8.6cm]{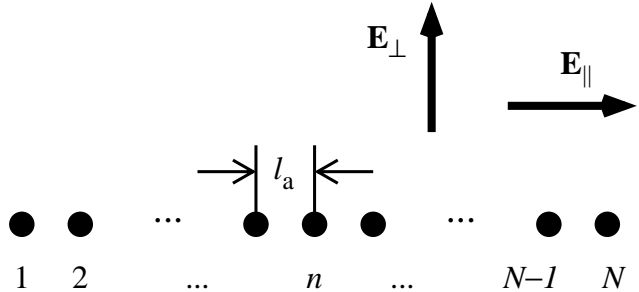}
\caption{%
1D array of atoms and laser light with different polarizations;
light is incident normally to the plane of the graph.}
\label{fig1}
\end{figure}

We consider here the most basic model of a
1D array of $N$ atoms lined up
along the $z$ axis, spaced by $l_a$,
and driven by a laser propagating normally to the array
and having an arbitrary polarization (Fig.~\ref{fig1}).
In the linear case, $|E_\mathrm{L}|^2 \ll E_\mathrm{sat}^2$,
when in (\ref{eq2.9}) $f_\mathrm{NL} = 1$,
it is sufficient to consider effects caused
by linearly polarized light with either
one of the two mutually orthogonal polarizations;
any other polarization (e.\,g.\ a circular one) can be treated
as a linear combination of those two.
In the case of a 1D array, natural choices
for these two basic configurations are:
\begin{itemize}
\item[(a)]
the incident field $\vect{E}_\mathrm{in}$ 
is parallel to the $z$ axis, 
$\vect{E}_\mathrm{in} \parallel \hat{\vect{e}}_z$
(and the dipoles line up ``head-to-tail''; we
will call it $\parallel$-configuration),
and 
\item[(b)]
$\vect{E}_\mathrm{in}$ is normal to the $z$ axis,
$\vect{E}_\mathrm{in} \perp \hat{\vect{e}}_z$
(``side-by-side'' lineup; $\perp$-configuration).
\end{itemize}
The general solution will be a
linear vectorial superposition of these two.
This choice of the basic configurations is dictated by the simplicity
of the resulting polarization of the local field. 
Indeed, in both cases, it follows that 
the polarization of local field
is parallel to that of the incident field,
$\vect{E}_\mathrm{L} \parallel \vect{E}_\mathrm{in}$, 
so we can use scalar equations for all fields.
Using the dimensionless notation 
$\vect{\mathcal{E}}_n 
= [E_\mathrm{L} (\vect{r}_n) / E_\mathrm{in}]_{(p)}$,
where $(p)$ denotes polarization,
$(p) =\: \parallel$ or $(p) =\: \perp$, 
and recalling that now $\vect{u} \parallel \hat{\vect{e}}_z$,
we write (\ref{eq2.9}) for both configuration as
\begin{eqnarray}
&&
\mathcal{E}_n + Q F_{(p)} \sum_{1 \le j \le N}^{j \ne n} 
    \frac{\mathcal{E}_j / 2}{|j - n|^3} = 1,
\quad\mbox{if}\quad  
n = 1,\ldots, N,
\nonumber
\\
&&
\mbox{and}\quad  \mathcal{E}_n = 0  \quad\mbox{otherwise},
\label{eq3.1}
\end{eqnarray}
where $F_{(p)}$ is a form-factor due to
polarization configuration:
$F_\parallel = 1$ and $F_\perp = - 1/2$.

If the 1D array is infinite, $N \to \infty$,
or sufficiently long, $N \gg 1$, it is instructive
to rewrite (\ref{eq3.1}) in the form
\begin{equation}
\mathcal{E}_n + S Q F_{(p)} \sum_{-\infty \le j \le \infty}^{j \ne n}
    \frac{\mathcal{E}_j / 2S}{|j - n|^3} = 1,      
\label{eq3.2}
\end{equation}
where $S = \sum_{j = 1}^\infty j^{-3} \approx 1.202057$.
The sums over $|j - n|^{-3}$ in (\ref{eq3.1}), (\ref{eq3.2}) converge
rather fast, hence $S - 1$ is not too large, see also 
Appendix~\ref{app} below.
Eqs.\ (\ref{eq3.1}) and (\ref{eq3.2}), the same as master 
equation (\ref{eq2.9}),
represent the case of \emph{fully-interacting arrays} (FIA),
whereby each atom ``talks'' to all the other
atoms in the array, which presents a challenge
to an analytical treatment.
Of course, a linear equation (\ref{eq3.1})
for $\mathcal{E}_n$ is solved analytically
using a standard linear algebra approach with matrices.
However, analyzing the results for large-size arrays, $N \gg 1$,
in particular analytically finding all the resonances
in $(\mathcal{E}_n)_\mathrm{max} (\delta)$,
can only be done by using numerical matrix solver,
even if we neglect the dissipation.

Thus, there is a need for a simple approximation
that would preserve most of the qualitative 
features of the phenomenon, yet could
be easily analyzed analytically.
This can be done by using the near-neighbor approximation (NNA),
similar to that of the Ising model of (anti)ferromagnetism,
in which the full sum in (\ref{eq3.1}) or (\ref{eq3.2}) 
is replaced by the sum over the nearest neighbors,
\begin{eqnarray}
&&
\mathcal{E}_n + \frac{Q F_{(p)}}{2} 
    (\mathcal{E}_{n - 1} + \mathcal{E}_{n + 1}) = 1,
\nonumber
\\
&&
\mathcal{E}_0 = \mathcal{E}_{N + 1} = 0.
\label{eq3.3}
\end{eqnarray}
In the ultimate two-atom case, $N = 2$ \cite{cit1}, 
the two approaches merge.
The further two-near-neighbors approximation (2-NNA)
and even three-near-neighbors approximation
are considered in Section~\ref{sec6} below.
We found, however, that
in general, a full summation (FIA) in (\ref{eq3.1}) on the one hand,
and NNA (\ref{eq3.3}) as well as 2-NNA on the other hand,
produce qualitatively similar results that 
differ by a factor of $O(1)$.

\begin{figure}
\includegraphics[width=8.6cm]{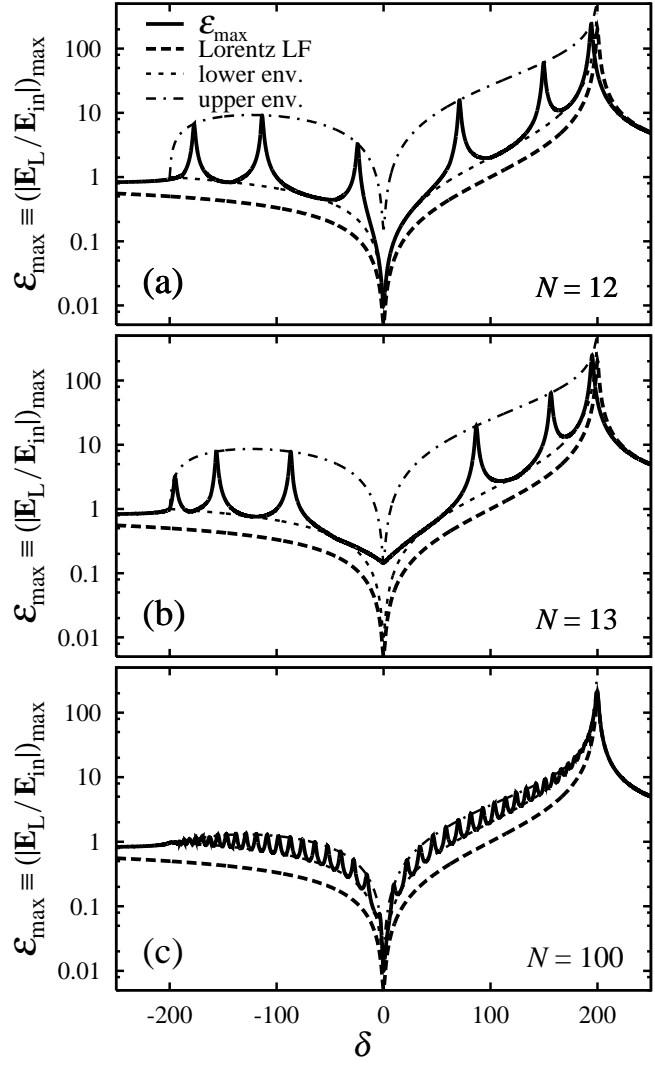}
\caption{%
Spectra of \emph{absolute}
maximum local field amplitudes near an atomic resonance 
($\delta = 0$) in the near-neighbor approximation with 
$\delta_\mathrm{LL} = 200$:
(a) locsiton resonances with 12 atoms in the array;
(b) the same with 13 (magic number) atoms in the array;
(c) merging and damping of locsiton resonances 
for a large number of atoms, $N = 100$.
The curves show the amplitudes of local fields for size-related resonances, 
local fields due to Lorentz--Lorenz theory,
and lower and upper amplitude envelopes of the resonances.}
\label{fig2}
\end{figure}

Since effects discussed here
are most pronounced in relatively small systems 
or in the small vicinity of perturbations
in large lattices, it is natural to stipulate
that the \emph{incident} field within the array
is uniform, unless otherwise is stated;
however, this condition can readily be arranged
even for an array larger than $\lambda$.
One of the solutions for the local field
(and atomic excitation) in the \emph{infinite} 1D array
(or a sufficiently long one, whereby we can neglect
edge effects) is also uniform. 
We will call it the ``Lorentz'' solution, $\bar{\mathcal{E}}$,
to be found from (\ref{eq3.2}) by setting
$\mathcal{E}_n = \mathcal{E}_j  = \bar{\mathcal{E}}$ as:
\begin{eqnarray}
&&
\bar{\mathcal{E}}_{(p)} = \frac{1}{1 +  Q S F_{(p)}} 
= \frac{1}{1 - [(\delta_\mathrm{LL})_{(p)} / (\delta  + i)]},
\nonumber
\\
&&
(\delta_\mathrm{LL})_{(p)} = - S Q_a F_{(p)},
\label{eq3.4}
\end{eqnarray}
where we introduced polarization-dependent 
parameter $\delta_\mathrm{LL}$,
which determines Lorentz--Lorenz
shift at $\delta = \delta_\mathrm{LL}$ (see below);
$|\delta_\mathrm{LL}|$ is a measure of 
the polarization-related strength of interaction.
Eq.\ (\ref{eq3.4}) may be viewed as a 1D counterpart of
the \emph{Lorentz--Lorenz} relation for local field;
notice, however, that the field $\bar{\mathcal{E}}$ is strongly
\emph{anisotropic} with respect to the polarization.
The spectral behavior of $|\bar{\mathcal{E}}|$
is depicted in Fig.~\ref{fig2} with thicker dashed curves 
in all the graphs.
If $Q_a \gg 1$, it shows, as one may expect,
a deep dip at the atomic resonance frequency, i.\,e.\ at $\delta = 0$,
\begin{equation}
|\bar{\mathcal{E}}_{(p)}|_\mathrm{min}^2 
= \frac{1}{1 +  \delta_\mathrm{LL}^2}
\label{eq3.5}
\end{equation}
(so that at the atomic resonance 
the local field is suppressed, as if 
it is pushed out of the  array),
and a strong new resonant peak appears at the shifted frequency
$\delta = \delta_\mathrm{LL}$ due to the Lorentz--Lorenz effect:
\begin{equation}
|\bar{\mathcal{E}}_{(p)}|_\mathrm{max}^2 = 1 + \delta_\mathrm{LL}^2 
= \frac{1}{|\bar{\mathcal{E}}_{(p)}|_\mathrm{min}^2},
\label{eq3.6}
\end{equation}
whose nature is essentially similar to e.\,g.\ Lorentz--Lorenz resonance
observed experimentally in alkali vapors \cite{cit6}.
However, in the 1D case considered here,
the Lorentz--Lorenz shift and even its sign are polarization-dependent;
in particular,
\begin{equation}
(\delta_\mathrm{LL})_\parallel = -  Q_a S,
\qquad
(\delta_\mathrm{LL})_\perp =  Q_a S / 2,
\label{eq3.7}
\end{equation}
i.\,e.\ the Lorentz--Lorenz resonance is red-shifted
for $\parallel$-polarization of laser, and blue-shifted
for $\perp$-polarization.
The Lorentz field $\bar{\mathcal{E}}$ and Lorentz--Lorenz shift
in the near-neighbor approximation, Eq.\ (\ref{eq3.3}), 
are determined by the same equations (\ref{eq3.4})--(\ref{eq3.7}),
where one has to set $S = 1$.

\section{Spatially-periodic and wave solutions (locsitons)}
\label{sec4}

We look now for solution of Eq.\ (\ref{eq3.2}) as the sum of \emph{uniform}
LF, $\bar{\mathcal{E}}$, Eq.\ (\ref{eq3.4}), and oscillating ansatz
$\Delta \mathcal{E} \propto \exp(\pm iqn)$,
where $q$ is an (unknown) wavenumber,
similarly, e.\,g.\ to the phonon theory \cite{cit9},
with the difference being that we have here
an excitation of bound electrons, and not atomic vibrations.
Essentially, the locsitons may be classified as \emph{Frenkel} 
excitons \cite{cit9}
because of their no-electron-exchange nature.

\begin{figure}
\includegraphics[width=8.6cm]{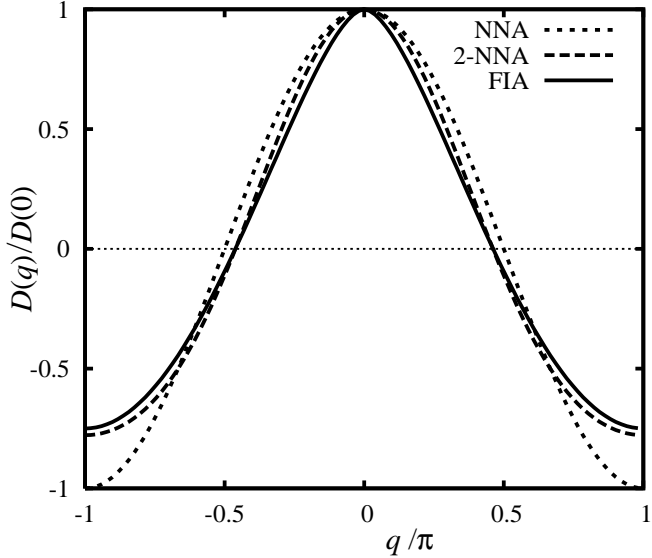}
\caption{%
Dispersion relations
for an infinite array and negligible losses,
i.\,e.\ normalized laser detuning $\delta / \delta_\mathrm{LL}$ 
\emph{vs} the normalized wavenumber, $q/\pi$.
Curves show the dispersion
for near-neighbor approximation, NNA (\ref{eq4.2}),
two-near-neighbors approximation, 2-NNA (\ref{eq6.6}),
and for fully interacting array (FIA),
whereby each atom interacts with all the other
atoms (\ref{eq4.1}), which coincides with the analytical 
fit (\ref{eqA.10}).}
\label{fig3}
\end{figure}

The wavenumbers $q$ are found \emph{via} the dispersion relation
\begin{equation}
D(q) \equiv \frac{1}{S} \sum_{n = 1}^\infty \frac{\cos(nq)}{n^3}
= \frac{\delta + i}{\delta_\mathrm{LL}}. 
\label{eq4.1}
\end{equation}
The behavior of $D(q)$ in the lossless
case, $\delta_\mathrm{LL}^2 > \delta^2 \gg 1$,
is depicted in Fig.~\ref{fig3} by the solid curve.
Within NNA, we have to set $S = 1$ and replace
the sum in (\ref{eq4.1}) by its first term:
\begin{equation}
D_\mathrm{NNA}(q) \equiv \cos q = \frac{\delta + i}{\delta_\mathrm{LL}},
\label{eq4.2}
\end{equation}
see Fig.~\ref{fig3}, fine-dashed curve.
Distinct oscillations emerge in the area
between the two edges of the locsiton band.
Their wavenumbers are determined from
(\ref{eq4.1}) or (\ref{eq4.2}) by neglecting the dissipation,
i.\,e.\ by assuming real $q$ and 
$\delta^2, \delta_\mathrm{LL}^2 \gg 1$. 
One of the band edges corresponds
to the maximum (and positive) $[D(q)]_\mathrm{max} = 1$ 
(same as for NNA, $[D_\mathrm{NNA}(q)]_\mathrm{max} = D(0) = 1$),
and thus to the Lorentz--Lorenz shift, 
$\delta  = \delta_\mathrm{LL}$.
The other, ``anti-Lorentz'', edge is at the opposite
side of the atomic resonance, and is determined
by the minimum (and negative) $D_\mathrm{min} = D(\pi)$.
Thus we have:
\begin{equation}
\frac{\delta_\mathrm{anti}}{\delta_\mathrm{LL}} = \frac{D(\pi)}{D(0)},
\label{eq4.3}
\end{equation}
which can be evaluated using an amazingly simple
relation for the sum (\ref{eq4.1}), which is
actually valid for a more general
sum and an arbitrary exponent $\rho > 1$:
\begin{equation}
\frac{D(\pi, \rho)}{D(0, \rho)} 
\equiv \frac{\displaystyle \sum_{n = 1}^\infty (-1)^n n^{-\rho}}
            {\displaystyle \sum_{n = 1}^\infty n^{-\rho}}
= -1 + \frac{1}{2^{\rho - 1}}.
\label{eq4.4}
\end{equation}
This can be readily proven by an appropriate rearrangement
of the terms in the sums in (\ref{eq4.4}), see below (\ref{eqA.11}).
In the case of $\rho = 3$ and infinite array, (\ref{eq4.1}),
we have $\delta_\mathrm{anti} / \delta_\mathrm{LL} = - 3/4$.
Hence, the locsiton band is determined as:
\begin{equation}
1 > \frac{\delta}{\delta_\mathrm{LL}} > - \frac{3}{4}
\qquad (\mbox{or }
\left| \frac{\delta}{\delta_\mathrm{LL}} \right| < 1 
\mbox{ within NNA}),
\label{eq4.5}
\end{equation}
with well developed locsitons at $\delta_\mathrm{LL}^2 \gg 1$.
Indeed, if the dissipation is neglected
and the wavenumbers $q$ are real, there is an infinite
number of solutions for them within the limits (\ref{eq4.5});
in this case the meaningful positive solutions,
within the first Brillouin zone, are $-\pi \le q < \pi$.
The plots of the functions in the l.\,h.\,p.\ of (\ref{eq4.1})
and (\ref{eq4.2}) are shown as $\delta / \delta_\mathrm{LL}$
vs $q$ in Fig.~\ref{fig3}.
Based on (\ref{eq4.5}), the total width of the locsiton
band in terms of $\delta$ is thus
$(7/4) |\delta_\mathrm{LL}|$, if one accounts for
the interactions of each atom with \emph{all} the 
rest of atoms in the infinite array,
whereas it is $2|\delta_\mathrm{LL}|$ 
in the near-neighbor approximation.

To gauge the dipole--dipole interaction
in the lattice, one can also introduce its Rabi-energy as:
\begin{equation}
\hbar \Omega_\mathrm{R} = \frac{\hbar Q_a}{T}
= \frac{4 |F_{(p)}| |d_a|^2}{\epsilon l_a^3} \ll \hbar \omega_0.
\label{eq4.6}
\end{equation}
It brings about a locsiton energy
band $\sim 2 \hbar \Omega_\mathrm{R} \gg \hbar \Gamma$
(if $\delta_\mathrm{LL}^2 \gg 1$) akin to those in solid-state \cite{cit9},
photonic crystals \cite{cit10}, and electronic band-pass filters.

\begin{figure*}
\includegraphics[width=16cm]{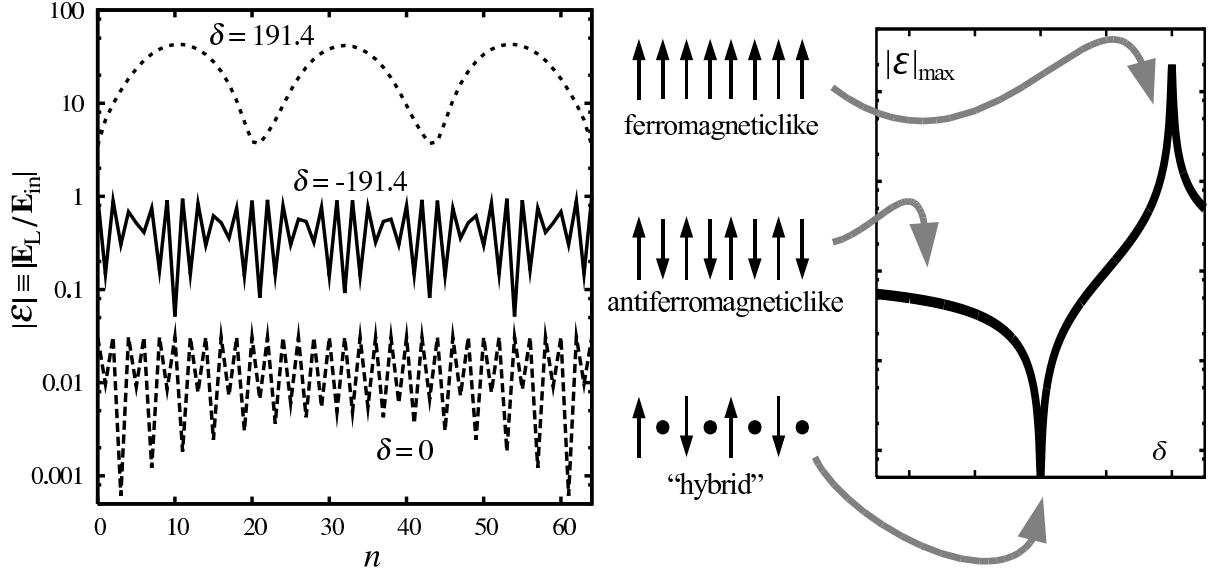}
\caption{%
Strata patterns of excitation and local field in finite arrays,
and their relations to the resonance tuning
in the case of 64 atoms and $\delta_\mathrm{LL} = 200$.
Curves and patterns show long-wave, 
ferromagneticlike excitation
near the Lorentz--Lorenz resonance
(top curve in the left plot and top pattern in the center),
counter-phase, antiferromagneticlike excitation
near the anti-Lorentz edge of the band
(middle curve and pattern), and
hybrid excitation near the point of atomic resonance
(bottom curve and pattern).
Note that all the curves in the left plot show 
\emph {absolute} normalized amplitudes of the local field.
Since the fields are in general complex,
their absolute amplitudes are positive,
so the near-zero points in the schematic
depiction of the ``hybrid'' mode
actually correspond to the lowest points of
the bottom curve in the left plot.
}
\label{fig4}
\end{figure*}

One of the most interesting effects
due to locsitons is a wide spectrum of the
standing waves, strata, formed by  them, see Fig.~\ref{fig4}.
As we already noted, they can range
from the very long ones, long-wave (LW) strata,
with the maximum spatial period being double
of the whole length of a 1D array,
similarly to the main mode of oscillation
in a violin string, to short-wave (SW) strata
whereby each dipole oscillates in 
counter-phase to each of its nearest neighbors.
The latter mode is the strongest manifestation of the fact
that the array of $N$ atoms is a discrete-element resonator,
akin a string of beads connected to each
other, with the beads capable of the same
kind of motion, when each individual bead is 
oscillating in counter-phase with its neighbor.
Because in both cases the excitation is of dynamic nature
and is due to external driving,
the mechanical analogy of SW locsiton modes
is more adequate than that of a static ferromagnetic
configuration \emph{vs} antiferromagnetic one;
see also Appendix~\ref{app} below.

It can be immediately found, e.\,g.\ from consideration
of the plots at Fig.~\ref{fig4}, upper curve, 
that the LW strata
emerge at the laser tuning very near to the
Lorentz--Lorenz resonance,
i.\,e.\ in the limit $1 - \delta / \delta_\mathrm{LL} \ll 1$.
In this case, in both FIA (\ref{eq4.1})
and NNA (\ref{eq4.2}), the wave number $q_\mathrm{LW}$ and 
the respective spatial wavelength $\Lambda_\mathrm{LW}$ are as:
\begin{equation}
q_\mathrm{LW} \approx \sqrt{1 - (\delta / \delta_\mathrm{LL} )^2},
\qquad
\Lambda_\mathrm{LW} = \frac{2 \pi l_a}{q_\mathrm{LW}}.
\label{eq4.7}
\end{equation}
In the infinite array, the 
longest $\Lambda$ is up to $2 \pi l_a \delta_\mathrm{LL}$,
when $\delta_\mathrm{LL}^2 - \delta^2 \sim 1$;
the locsitons with longer wavelength get significantly suppressed.

The opposite limit, or the anti-Lorentz side of the locsiton band, 
\begin{equation}
\frac{3}{4} + \frac{\delta}{\delta_\mathrm{LL}} \ll 1
\qquad
(\mbox{for NNA it is}\quad
1 + \frac{\delta}{\delta_\mathrm{LL}} \ll 1 ),
\label{eq4.8}
\end{equation}
defines SW locsitons, with 
\begin{equation}
q_\mathrm{SW} \sim \pi,
\qquad
\Lambda_\mathrm{SW} / 2 \sim l_a,
\label{eq4.9}
\end{equation}
i.\,e.\ $\Lambda_\mathrm{SW} / 2$
is the finest grain of locsiton structure,
as one would expect from the ``bead string'' analogy.
However, an incommensurability, i.\,e.\ 
mismatch between $\Lambda_\mathrm{SW} / 2$
and the lattice spacing $l_a$,
whose ratio is in general an \emph{irrational}
number, results in a strong spatial modulation of the SW,
giving rise to a \emph{coarse} LW-like structure.

In the NNA, this coarse structure of SW mode
has the half-period roughly the same as for a
pure LW mode, $\sim \Lambda_\mathrm{LW}$.
This can be readily understood in terms of
``beating'' between the locsiton wavelength, $2\pi/q$,
and  the spatial scale of the discrete structure of the system,
which is the normalized spacing between atoms, 1.
Indeed, near the anti-Lorentz
edge, $\delta + \delta_\mathrm{LL} \ll \delta_\mathrm{LL}$, 
we can write for a SW wavenumber: $q_\mathrm{SW} = \pi - \Delta q$,
with 
$|\Delta q| \approx \sqrt{1 - (\delta / \delta_\mathrm{LL} )^2} 
= q_\mathrm{LW}$,
and find the spatial oscillations as
\begin{eqnarray}
\mathcal{E}_n &\propto& \cos(n q_\mathrm{SW}) 
= \cos(n\pi - n\Delta q)
\nonumber
\\ 
&=& (-1)^n \cos(n\Delta q),
\label{eq4.10}
\end{eqnarray}
with $\Delta q \approx q_\mathrm{LW}$,
which shows alternating, counter-phase motion of the neighboring
atoms, $(-1)^n$, modulated by a slow envelope, 
$\cos(n q_\mathrm{LW})$.
Both the fine grain and the coarse modulation
may be well pronounced (Fig.~\ref{fig4}, middle curve).
At $q = \pi$, the LW and SW periods converge to
the same scale, $4 l_a$, see (\ref{eq4.12}) below.
Using the phonon analogy, the LW locsitons may be
viewed to a certain extent
as counterparts to acoustic, and SW -- to optical phonons.

Between those limiting points -- LW locsitons
at the Lorentz end of the locsiton spectrum, 
$\delta \sim \delta_\mathrm{LL}$, and
SW locsitons at the anti-Lorentz end, 
$\delta \sim -\delta_\mathrm{LL}$, --
there are all kinds of locsitons making
chaotic looking strata (due to the above mentioned
incommensurability, i.\,e.\
irrational ratio between a locsiton wavelength, $\Lambda$, 
and the array spacing, $l_a$).
However, same as in the chaotic motion,
there are small islands of well ordered 
wave patterns, located at the spectral points
where the ratio $\Lambda / l_a$ (or $q/\pi$)
is a rational number, provided that
the system is at a \emph{resonance}, 
see (\ref{eq5.2}) and (\ref{eq5.3}) below.
One can think of them as sort of hybrids of ferromagneticlike 
and antiferromagneticlike behavior.
Indeed, the purely antiferromagneticlike
SW locsiton at $q = \pi$
is formed by the atoms with alternating polarizations,
\begin{eqnarray}
&&
\cdots\:\: \uparrow\:\: \downarrow\:\: \uparrow\:\: \downarrow\:\: 
\uparrow\:\: \downarrow\: \cdots
\quad\mbox{for $\perp$ polarization, and}
\nonumber
\\
&&
\cdots\: \rightarrow\: \leftarrow\: \rightarrow\: \leftarrow\: \rightarrow\: 
\leftarrow\: \cdots
\quad\mbox{for $\parallel$ polarization}.
\label{eq4.11}
\end{eqnarray}
This happens  at $\delta \sim -\delta_\mathrm{LL}$
within NNA and $\delta \sim - (3/4) \delta_\mathrm{LL}$ for FIA.
The simplest hybrid pattern  is formed at 
$q = \pi/2$ [which is at $\delta = 0$ within NNA,
and $\delta / \delta_\mathrm{LL} = -3/32$ for a fully
interacting array (\ref{eq4.1})],
whereby each second atom is \emph{non-excited}, while the
other atoms alternate their polarization:
\begin{eqnarray}
&&
\cdots\: \uparrow \circ \downarrow \circ \uparrow \circ
\downarrow \circ \uparrow \circ \downarrow \cdots
\quad\mbox{for $\perp$ polarization, and}
\nonumber
\\
&&
\cdots \rightarrow \circ \leftarrow \circ \rightarrow \circ
\leftarrow \circ \rightarrow \circ \leftarrow \cdots
\label{eq4.12}
\\
&&
\qquad\qquad\qquad\qquad\qquad\qquad\qquad
\mbox{for $\parallel$ polarization}.
\nonumber
\end{eqnarray}
Examples of other simplest hybrid states within NNA are:
$q = \pi/3$, \ $q = \pi/4$, etc.;
in general, periodic patterns exist
if $q/\pi$ is a rational number,
i.\,e.\ for all the eigen-resonances in finite
1D arrays within NNA, see Eq.\ (\ref{eq5.3}) below.

Finally, it is worth noting that the strata, albeit fast-decaying,
exist even \emph{beyond} the locsiton band, (\ref{eq4.5}). 
They are not, however, propagating locsitons,
and their amplitudes exponentially decay with the distance. 
Consider the simplest case, NNA,
with the losses negligibly small,
$\delta^2 > \delta_\mathrm{LL}^2 \gg 1$ in (\ref{eq4.2}).
In this case, $\cos^2 q > 1$ in (\ref{eq4.2}),
which indicates that the wavenumber $q$ must be complex.
Indeed, writing $q = -i\chi$ for the ``Lorentz''
end of the band, $\delta / \delta_\mathrm{L} > 1$,
and $q = -i\chi + \pi$ for the anti-Lorentz end,
$\delta / \delta_\mathrm{L} < -1$ (see also Section~\ref{sec6} below),
we have the solution for $\chi$ as 
\begin{equation}
\chi = \ln \left[ |\delta / \delta_\mathrm{LL}|
    \pm \sqrt{(\delta / \delta_\mathrm{LL})^2 - 1} \right]
\label{eq4.13}
\end{equation}
and for the local fields as
\begin{equation}
(\mathcal{E}_n)_\mathrm{LL} \propto e^{\chi n}
    \qquad\mbox{and}\qquad
(\mathcal{E}_n)_\mathrm{antiLL} \propto e^{\chi n} (-1)^n
\label{eq4.14}
\end{equation}
for the Lorentz and anti-Lorentz ends of the band, respectively.
Here, in the case of a semi-infinite array,
the sign in (\ref{eq4.13}) has to be chosen such
that the amplitude of the field vanishes at $|n| \to \infty$.
These modes can be viewed as \emph{evanescent waves\slash{}locsitons},
see also below, Section~\ref{sec6}.
One can note though that they still bear the signature
of the respective locsitons on either side of the band:
long-wave, almost synchronous oscillations
on the Lorentz end, and short-wave, phase-alternating 
oscillations on the anti-Lorentz end.
In general, the same patterns hold in the FIA case.

\section{Quantization of locsitons and field resonances in finite 1D arrays}
\label{sec5}

Due to boundary conditions in (\ref{eq3.1}) [or (\ref{eq3.3}) within NNA],
the array of $N$ atoms is a discrete-element resonator.
This should result in locsiton quantization 
within the locsiton band (\ref{eq4.5}) and corresponding
size-related resonances of the local field.
In a 1D array with $N$ atoms,
we have $N$ coupled oscillators with the same
individual atomic resonance frequency,
and therefore, one should expect the original
atomic line to be split into $N$ lines at most,
with the collective broadened band (\ref{eq4.5}) 
being replaced with those lines.
Of course, when the dissipation (or finite linewidth
of each individual line)
is taken into account, those split lines
will merge into one continuous locsiton band (\ref{eq4.5})
if the array is sufficiently large,
\begin{equation}
N > \frac{7}{4} |\delta_\mathrm{LL}|
\qquad  (\mbox{or } >2|\delta_\mathrm{LL}| \mbox{ within  NNA}).
\label{eq5.1}
\end{equation}

The simplest result for the resonant line positions
is obtained within NNA.
Using the boundary condition (\ref{eq3.3}), i.\,e.\
$\mathcal{E}_0 = \mathcal{E}_{N + 1} = 0$,
we find that the longest locsiton half-wave,
corresponding to the fundamental mode, is
\begin{equation}
\Lambda_1 / 2 = (N + 1) l_a,
\qquad
q_1 = \pi / (N + 1).
\label{eq5.2}
\end{equation}
Thus, $\Lambda_1 / 2$ is the distance 
between nodes where LF zeros out,
whereas the wavelengths $\Lambda_k$
of eigen-locsitons and their eigen-frequencies
$\delta_k$, with the quantum number 
$1 \le k \le N$, are respectively
\begin{eqnarray}
\Lambda_k &=& \Lambda_1 / k,
    \qquad
\delta_k = \delta_\mathrm{LL} \cos q_k,
\nonumber
\\
q_k &=& \pi k / (N + 1).
\label{eq5.3}
\end{eqnarray}
[Note that the first Eq.~(7) in \cite{cit1}, which corresponds to the
second Eq.~(\ref{eq5.3}) here, contained a typo (an extraneous $\pi$ in
the cos argument) which we corrected here.]
From these, only the resonances with odd $k$
will be realized for a symmetric driving laser profile,
in particular, the uniform one,
$E_\mathrm{in} = \mathrm{const}$ (which is the most common case here),
and with even $k$ -- for an anti-symmetric one, say,
$E_\mathrm{in} = \mathrm{const} \cdot (N + 1 - 2n) / (N - 1)$,
where $n$ is the sequence number of an atom in the array.
In all other cases, a full set of $N$
resonances will be realized.

In essence, the size-related locsiton resonances
in discrete arrays are, to a limited extent, similar
to any eigen-resonances in regular continuous, i.\,e.\
non-discrete, 1D system.
Examples can be found both in classical
setting, e.\,g.\ a violin string, a Fabri--Perot
resonator (as e.\,g.\ in a laser), and in quantum mechanics, from
the resonances in a quantum well with infinitely-high walls, to
electron gas in a finite layer \cite{cit11}, 
electrons in long molecules \cite{cit12}, etc.
The major difference here is that 
the number of eigenmodes, or resonances, in an
array with $N$ elements is limited to $N$, in
contrast to the theoretically infinite number of
eigenmodes in continuous finite 1D systems.

The resonances for uniform driving within NNA
are shown in Fig.~\ref{fig2} for $\delta_\mathrm{LL} = 200$
in the case of $N = 12$ [Fig.~\ref{fig2}(a)], 
$N = 13$ [Fig.~\ref{fig2}(b)],
and $N = 100$ [Fig.~\ref{fig2}(c)].
One can readily find out that
the lower amplitude envelope is
\begin{equation}
\mathcal{E}_\mathrm{min} (\delta) \approx 2|\bar{\mathcal{E}}|,
\label{eq5.4}
\end{equation}
while the upper envelope of the resonant peaks within NNA
for a uniform driving is
\begin{equation}
\mathcal{E}_\mathrm{max}
\begin{cases}
\approx |\bar{\mathcal{E}}| (n_{\delta} + n_{\delta}^{-1}),
\quad\mbox{if}\quad  n_\delta \le 1,
\\
= 2|\bar{\mathcal{E}}|
\quad\mbox{otherwise},
\end{cases}
\label{eq5.5}
\end{equation}
where $n_\delta = (N+1) / (2 \sqrt{\delta_\mathrm{LL}^2 - \delta^2})$.
As $N$ increases, the resonances merge
and are suppressed at $N = |\delta_\mathrm{LL}| O(1)$,
see e.\,g.\ Fig.~\ref{fig2}(c).
However, even then $\mathcal{E}_\mathrm{min}$
still exceeds the uniform field
$|\bar{\mathcal{E}}|$ (\ref{eq3.4}) by a factor of 2.
For $N = 3k - 1$ ($k$ is a natural number),
LF amplitude dips below the lower envelope 
$|\mathcal{E}_\mathrm{max}|_\mathrm{low}$
at $\delta = -\delta_\mathrm{LL} / 2$.
At that frequency, within NNA, $\cos q' = -0.5$,
$q' = 2\pi / 3$, and the SW period $\Lambda = 3 l_a$
is an integer of the atomic
spacing, so only fine SW structure remains,
resulting in an \emph{anti-resonance} and
in the strongest inhibition of the locsiton.

\section{1D arrays beyond the near-neighbor approximation}
\label{sec6}

As we mentioned above,
while the quantization of locsitons in finite arrays
can be readily analyzed analytically within NNA,
see the previous Sections \ref{sec3} and~\ref{sec5},
the situation with fully-interacting arrays (FIA)
presents a challenge for an analytical treatment.

Let us briefly outline general analytical
and numerical results obtained so far.
\begin{itemize}
\item
A FIA locsiton band is not symmetric
with respect to the atomic resonance, $\delta = 0$;
it is shorter by the factor $3/4$
on the anti-Lorentz side, see (\ref{eq4.5}),
as opposite to the NNA.
\item
Respectively, FIA resonances  are grouped tighter
on the anti-Lorentz side of the band,
albeit their number is the same as for NNA.
Near the Lorentz side of the band, the 
NNA-predicted resonances coincide more closely with
those obtained by FIA numerical calculations.
\end{itemize}

The major source of these effects is the first factor, i.\,e.\
a strongly asymmetric (with respect to the 
detuning frequency, $\delta$)
shape of the dispersion relation, Fig.~\ref{fig3}.
A more detailed mathematical consideration
of this problem, including a very good \emph{analytical}
fit for the dispersion relation is found in 
Appendix~\ref{app} below.
Less significant, although interesting as far as the
eigenmodes of 1D arrays are concerned, is the fact that
simple NNA eigen-wavenumbers (\ref{eq5.3}),
$q_k = \pi k / (N + 1)$, obtained
based on the NNA boundary conditions
$\mathcal{E}_0 = \mathcal{E}_{N + 1} = 0$ [Eq.\ (\ref{eq3.3})]
are not exact anymore.
One has to use now more extended, ``beyond-the-boundary'' 
conditions (\ref{eq3.1}), which are the signature of FIA, whereby 
\begin{equation}
\mathcal{E}_n = 0
\quad\mbox{for \emph{all}}\quad  n < 1
\quad\mbox{and}\quad n > N,
\label{eq6.1}
\end{equation}
instead of just two end-points in NNA.
To explore the problem, let us simplify it first 
by considering only non-dissipating atoms, 
$\delta_\mathrm{LL}^2 \ge \delta^2 \gg 1$,
and rewrite the full-interaction dispersion relation
(\ref{eq4.1}) for this case as:
\begin{equation}
\frac{1}{S} \sum_{n =1}^\infty \frac{\cos(nq)}{n^3}
= D(q) \equiv \frac{\delta}{\delta_\mathrm{LL}},
\label{eq6.2}
\end{equation}
assuming $D(q)$ real.
The equation for the field is written then as:
\begin{equation}
\mathcal{E}_n - \frac{1}{D} \sum_{-\infty \le j \le \infty}^{j \ne n}
    \frac{\mathcal{E}_j / 2 S}{|j - n|^3} = 1,     
\label{eq6.3}
\end{equation}
with the boundary conditions (\ref{eq6.1}).

The plot $D(q)$ for real $q$'s 
due to the dispersion relation (\ref{eq6.2})
is shown in Fig.~\ref{fig3} with the solid curve.
The new qualitative difference now between FIA, (\ref{eq6.2}),
and the NNA case, (\ref{eq4.2}) without dissipation,
i.\,e.\ $\cos q = D$, is as follows.
With $D^2 \le 1$, the NNA equation has \emph{only real}
solutions for $q$, whereas Eq.\ (\ref{eq6.2}), even within the locsiton
band, $-3/4 \le D \le 1$,
aside from real solution for $q \equiv \tilde{q}$,
has, as one can show, an \emph{infinite} number of 
\emph{complex} solutions, $q$, for \emph{each} single
real solution $\tilde{q}$ (i.\,e.\ for the same $D$).
All of them, in addition to fast spatial oscillation
terms, have a rapidly rising\slash{}falling exponential factor.
These exponential modes are negligibly small almost over
the entire array length, if $N \gg 1$,
and they need to be accounted for only very near
the end-points of the array,
where they are instrumental in zeroing out
the field and excitation at the points 
$n < 1$ and $n > N$.

Let us illustrate the formation of those exponential
(or \emph{evanescent}) modes and their role in boundary conditions
for the \emph{two}-near-neighbors approximation,
\emph{2-NNA}, whereby the field Eq.\ (\ref{eq6.3}) becomes:
\begin{equation}
\mathcal{E}_n - \frac{9}{4D}
\left[ (\mathcal{E}_{n - 1} + \mathcal{E}_{n + 1})
 + \frac{1}{8}(\mathcal{E}_{n - 2} + \mathcal{E}_{n + 2}) \right] = 1,
\label{eq6.4}
\end{equation}
with the boundary conditions for \emph{two} pairs
of end-points: 
\begin{equation}
\mathcal{E}_n = 0
    \quad\mbox{at}\quad  n = 0, -1
    \quad\mbox{and}\quad  n = N + 1,\, N + 2.
\label{eq6.5}
\end{equation}
The dispersion relations approximating (\ref{eq6.2}) 
will read now as:
\begin{equation}
\frac{8}{9} \left[\cos q + \frac{\cos(2q)}{8} \right] = D(q),
\label{eq6.6}
\end{equation}
see Fig.~\ref{fig3}, long-dashed curve,
and the locsiton band is determined by 
\begin{equation}
-7/9 \le D(q) \le 1.
\label{eq6.7}
\end{equation}
The real solutions $\tilde{q}$ of (\ref{eq6.6}) are those 
for which $\cos^2 \tilde{q} \le 1$;
they give rise to strata modes of Eq.\ (\ref{eq6.4}),
\begin{equation}
\mathcal{E}_n \propto \exp (\pm i\tilde{q} n).
\label{eq6.8}
\end{equation}
However, having in mind that
$\cos(2q) = 2 \cos^2 q - 1$,
one can readily see that  (\ref{eq6.6})
has also solutions with $\cos^2 q > 1$,
i.\,e.\ those that correspond to exponential
modes, with complex $q = q_\mathrm{evn2}$.
Introducing for those modes
\begin{equation}
q_\mathrm{evn2} = -i\chi + \pi
\label{eq6.9}
\end{equation}
with real $\chi$,
we obtain from (\ref{eq6.6}) that for each given real $\tilde{q}$ 
the exponent $\chi$ is determined by:
\begin{equation}
\cosh \chi = 4 + \cos \tilde{q},
\label{eq6.10}
\end{equation}
and the respective exponential mode is
\begin{equation}
(\mathcal{E}_n)_2 = e^{\pm (\chi + i\pi) n} = e^{\pm \chi n} (-1)^n.
\label{eq6.11}
\end{equation}
Thus, these 2-NNA modes are antiferromagneticlike
strata, modulated by fast exponents.
Indeed, since $3 \le \cosh \chi \le 5$,
we have $1.76 < \chi < 2.3$.

Since the exponential, or evanescent, modes have so short
``tails'', they produce relatively small
correction for the respective eigen-wavelengths, $\Lambda_n$,
and for $q_n$, compared to the NNA oscillatory modes,
(\ref{eq5.2}), (\ref{eq5.3}), so that we can look for the 
$\Lambda_n$ corrected for 2-NNA 
at the points of resonances as:
\begin{eqnarray}
\frac{\Lambda_k}{2 l_a} &=& \frac{N + 1 + \Delta_k}{k}, 
    \qquad
\Delta_k = O(1),
\nonumber
\\
\tilde{q}_k &=& \pi \frac{2 l_a}{\Lambda_k},
\label{eq6.12}
\end{eqnarray}
with the correction $\Delta_k > 0$,
similarly to, e.\,g., oscillations
in a violin string with a ``soft'' suspension at its ends.
To find $\Delta_k$ of the $k$-th resonance, 
we seek a solution for $\mathcal{E}_n$
as a sum of two modes: oscillatory and exponential ones.
Then, for symmetric modes, i.\,e.\ with odd $k$,
we have a full solution written as
\begin{eqnarray}
(\mathcal{E}_n)_\mathrm{odd} 
&=& \cos(\tilde{q}_k \bar{n}) + C \cosh[\chi (\tilde{q}_k) \bar{n}] (-1)^n,
\nonumber
\\
&&\mbox{with}\quad
\bar{n} = n - \frac{N + 1}{2},
\label{eq6.13}
\end{eqnarray}
where $C$ is a constant.
For anti-symmetric modes, with even $k$, we have:
\begin{equation}
(\mathcal{E}_n)_\mathrm{even} 
= \sin(\tilde{q}_k \bar{n}) + C \sinh[\chi (\tilde{q}_k) \bar{n}] (-1)^n.
\label{eq6.14}
\end{equation}
Using now conditions (\ref{eq6.5}), we can write for the points
$n = -1$ and $n = 0$, respectively, the following equations
for the symmetric modes (\ref{eq6.13}):
\begin{eqnarray}
(\mathcal{E}_{-1})_\mathrm{odd} 
&=& \cos \left( \frac{\pi k}{2} \frac{N + 3}{N + 1 + \Delta_k} \right)
\nonumber
\\
&&{}- C \cosh \left( \chi \frac{N + 3}{2} \right) = 0,
\label{eq6.15}
\\
(\mathcal{E}_0)_\mathrm{odd} 
&=& \cos \left( \frac{\pi k}{2} \frac{N + 1}{N + 1 + \Delta_k} \right)
\nonumber
\\
&&{}+ C \cosh \left( \chi \frac{N + 1}{2} \right) = 0.
\label{eq6.16}
\end{eqnarray}
For the anti-symmetric modes (\ref{eq6.14}),
one has to replace $\cos$ and $\cosh$ functions
in (\ref{eq6.15}), (\ref{eq6.16}) with $\sin$ and $\sinh$
functions, respectively.
From these two equations we can compute $\Delta_k$ and $C$.
Indeed, approximating 
$\cosh(\xi) \approx \sinh(\xi) \approx e^\xi / 2$ 
in (\ref{eq6.15}) and (\ref{eq6.16}) or in the respective
equations for anti-symmetric modes,
since $\chi (N + 1) \gg 1$,
we obtain two equations for $\Delta_k$ and $C$,
which are readily solved for $\Delta_k$ and $C$
for \emph{both} symmetric and anti-symmetric modes as:
\begin{eqnarray}
\Delta_k &=& \frac{2}{q_\mathrm{NNA}} 
\tan^{-1} \left[ \frac{\sin(q_\mathrm{NNA})}{e^\chi 
                 + \cos(q_\mathrm{NNA})} \right],
\label{eq6.18}
\\
&&\mbox{with}\quad    
q_\mathrm{NNA} = \frac{\pi k}{N + 1}
\nonumber
\\
&&\mbox{and}\quad
C = q_\mathrm{NNA} \Delta_k (-1)^{\bar{k}/2} e^{-\chi(N+1)},
\label{eq6.19}
\end{eqnarray}
where from (\ref{eq6.10}), 
$\chi \approx \cosh^{-1} [4 + \cos(q_\mathrm{NNA})]$,
while $\bar{k} = k + 1$ for symmetric modes,
and $\bar{k} = k$ for the anti-symmetric ones.
It is worth noting that the boundary conditions 
(\ref{eq6.5}) at  $n = N + 1, N + 2$
are now satisfied automatically because of our choice
of the coordinate $\bar{n}$ in (\ref{eq6.13}) and (\ref{eq6.14}).
For LW resonances, $k \ll N$, (\ref{eq6.18}) reduces to
\begin{equation}
\Delta_k \approx \frac{2}{e^\chi + 1}
\approx \Delta_\mathrm{max} \approx \frac{2}{11},
\label{eq6.20}
\end{equation}
which is the maximum magnitude of $\Delta_k$,
whereas for SW resonances, $N + 1 - k \ll N + 1$, (\ref{eq6.18}) reduces to
\begin{equation}
\Delta_k \approx \frac{2}{e^\chi - 1} \left( \frac{N + 1}{k} - 1 \right)
\approx \frac{2}{5} \left( \frac{N + 1}{k} - 1 \right),
\label{eq6.21}
\end{equation}
which is substantially smaller than the LW correction (\ref{eq6.20}).
One can readily see that $\Delta_k$
is a monotonically decreasing function of $k$.
In the middle of the locsiton band, $k \sim (N + 1)/2$,
we have
\begin{equation}
\Delta_k \approx \frac{4}{\pi e^\chi} \approx \frac{1}{2\pi}.
\label{eq6.22}
\end{equation}
Interestingly, a fairly good
fit to (\ref{eq6.18}) is provided by a much simpler formula:
\begin{equation}
\Delta_k \approx \Delta_\mathrm{max} 
    \left[ 1 - \left(\frac{k}{N + 1}\right)^3 \right].
\label{eq6.23}
\end{equation}
Now, once the correction $\Delta_k$ is found, 
one can substitute $q = \tilde{q}_k = \pi k / (N + 1 + \Delta_k)$,
(\ref{eq6.12}), into the dispersion relation (\ref{eq6.6}) and
calculate the respective frequency
detuning for the $k$-th resonance, 
$D_k = \delta_k / \delta_\mathrm{L}$,
with $k = 1$ being the closest 
to the Lorentz--Lorenz resonance,
i.\,e.\ a LW mode, and $k = N$ closest to the
anti-Lorentz edge of the locsiton band, a SW mode.

A way to extend the 2-NNA approximation to
a full-array interaction is to apply the result
(\ref{eq6.18}) for the correction of the eigen-wavenumber, $\tilde{q}_k$,
but use it now in the full-blown
dispersion relation (\ref{eq6.2}), instead of 
the 2-NNA relation (\ref{eq6.6}),
to calculate the resonance detuning $\delta_k$.
The other avenue, of course, 
is to seek for higher order approximations.
For example, one can take into account two edge sets
of \emph{3 points} each, i.\,e.\ similarly to (\ref{eq6.5}), 
stipulate 3-NNA:
\begin{eqnarray}
\mathcal{E}_n &=& 0
\quad\mbox{at}\quad  n = 0, -1, -2
\nonumber
\\
&&
\mbox{and}\quad  n = N + 1,\, N + 2,\, N + 3.
\label{eq6.24}
\end{eqnarray}
Following the same route as for 2-NNA case,
we obtain now, similarly to (\ref{eq6.10}), a \emph{second order}
equation for the \emph{exponential eigenmode} 
complex wavenumber, $q = q_\mathrm{evn3}$, for any given
oscillation wavenumber, $\tilde{q}$, as
\begin{eqnarray}
&&
16 \cos^2 q + \cos q\, (16 \cos\tilde{q} + 27)
\nonumber
\\
&&\quad{}
+ (104 + 16 \cos^2 \tilde{q} + 27 \cos\tilde{q}) = 0
\label{eq6.25}
\end{eqnarray}
[we leave it untransformed to $\cosh\chi$, unlike
(\ref{eq6.9}), (\ref{eq6.10}), since in 3-NNA case the solution 
becomes more complicated than (\ref{eq6.11}), and that transformation 
does not simplify the problem.]
The higher-order equation for the \emph{exponential eigenmodes}
ensue the choice of a higher number of 
end-points in the boundary conditions.

It has to be noted, however, that for
specifying parameters of oscillating eigenmodes, 
the increase of the precision by accounting for higher-order 
exponential modes is of very limited, if purely academic, significance.
Essentially, for large arrays, $N \gg 1$,
the combination of the NNA for predicting the wavenumbers
$\tilde{q}$, (\ref{eq6.12}) or (\ref{eq5.3}), 
with these $\tilde{q}$'s used then
in (\ref{eq6.2}) to calculate the resonance eigen-frequencies $\delta$,
does already a good job.
A further step in increasing the precision provided by the 2-NNA
corrections to eigen-wavenumbers, is to again use
the corrected $\tilde{q}$'s in (\ref{eq6.2}) for the same purpose,
and it is more than sufficient.
A special case is a small array, e.\,g.\ $N = 2, 3, 4$,
whereby it is actually preferable  to simply 
solve the general 1D-array equations analytically,
similarly to \cite{cit1} in the case of $N = 2$.

\section{Magic numbers}
\label{sec7}

A fundamental effect of self-induced
cancellation of local-field suppression
emerges near the atomic resonance, $\delta = 0$,
at certain ``magic'' numbers $N$.
If $\delta_\mathrm{LL}^2 \gg 1$,
the \emph{uniform} (Lorentz--Lorenz) LF (\ref{eq3.4})
at $\delta  = 0$ is very small,
$|\bar{\mathcal{E}}| \sim |\delta_\mathrm{LL}|^{-1}$ 
[Eq.\ (\ref{eq3.5})].
However, in the near-neighbor approximation, if 
\begin{equation}
N = k m_\mathrm{mag} +1,
\qquad  m = 1, 2, 3, \ldots,
\label{eq7.1}
\end{equation}
where $m_\mathrm{mag} = 4$ is a ``magic'' number within NNA,
locsitons at $\delta = 0$ show canceled LF suppression
at some atoms.
The highest cancellation is attained at $\delta = 0$ and $N = 5$, 
with the atomic dipoles lining up as in (\ref{eq4.12}),
whereby the LF amplitude at odd-numbered atoms is at maximum,
$|\mathcal{E}_\mathrm{max}| \approx 1/3$, 
and the enhancement 
$|\mathcal{E}_\mathrm{max}/\bar{\mathcal{E}}|_\mathrm{enh}^2 
\approx \delta_\mathrm{LL}^2 / 9$
could be a few orders of magnitude;
the LF at the two other atoms almost zeros out.

The self-induced cancellation effect 
is produced by a standing 
wave with the nodes at atoms with even numbers.
This results in a ``virtual'' size-related resonance at $\delta \to 0$
(i.\,e.\ at the exact atomic resonance), 
which manifests itself in the enhancement
(the resonant peak transpires in
$|\mathcal{E}_\mathrm{max}/\bar{\mathcal{E}}|$ \emph{vs} $\delta$).
Thus, \emph{the nature of magic numbers is
the coincidence of the atomic resonance
with one of the size-related locsiton resonances}.

The effect holds also for
the interaction of each atom with \emph{all} other atoms (\ref{eq3.1});
the magic number in (\ref{eq7.1})
takes on a ``devilish'' likeness here: $m_\mathrm{mag} = 13$.
It is due to the fact that
for $\delta = 0$ and $\delta_\mathrm{LL} \gg 1$,
the first root $q'$ of Eq.\ (\ref{eq4.1}) with zero r.\,h.\,s.,
\begin{equation}
\sum_{n = 1}^\infty \frac{\cos(nq')}{n^3} = 0,
\label{eq7.2}
\end{equation}
has its property of
$q' / \pi$ almost coinciding with a rational number,
$q' / \pi \approx$ $6/13$ \
($13 q'/ 6\pi  = 1.00026\ldots$),
see Appendix~\ref{app} below.
The locsiton wavelength is 
$\Lambda = 2\pi/q' = (13/3)l_a$, and the lowest
integer of $\Lambda/2$
to exactly match an integer of $l_a$
is $13\,l_a$, which requires 14 atoms.
We have now $|\mathcal{E}_\mathrm{max}| \approx 2/15$,
with enhancement $\sim 4 \delta_\mathrm{LL}^2 / 225$.

As we have shown in \cite{cit1}, 2D lattices can also
exhibit ``\emph{magic shapes}'' with similar
properties; the simplest one within NNA
is a six-point star with an atom at its center,
thus making the total number of atoms again $N = 13$.
More details on magic shapes for 2D lattices
will be discussed by us elsewhere.

\section{Traveling locsiton waves: velocity and penetration depth}
\label{sec8}

If the locsitons are waves, can they be excited
outside the driving field area?
and thus travel away from the Lorentz--Lorenz
uniform local field area?
how far can they travel before being extinguished?
how fast do they propagate?
having in mind their dissipation,
what is the size of a finite array 
to have well pronounced standing waves
and size-related resonances?

Of course, the locsitons can propagate
in an array even in the areas unaccessible
for the external (laser) field.
If the spatial profile gradient of the driving wave
is large enough, a LF excitation can be found
beyond the driving field area;
the terminological irony here is that the \emph{local}-field phenomenon
is due to a \emph{nonlocal} interaction, and
the locsitons can propagate away from
their origination point.
This may happen when the external laser field
is non-uniform and has a large gradient,
e.\,g.\ when one entirely screens out a part of the array
by imposing a ``sharp knife'' over the array.

Even simpler and more transparent case
is when only an end-point of a semi-infinite 
1D array [say, with $n = 1$ in
(\ref{eq3.1})] is illuminated by a laser field \emph{via} a pin-hole.
Eq.\ (\ref{eq3.1}) has then non-zero (unity) r.\,h.\,p.\
for $n = 1$ only, and the r.\,h.\,p.\ zeroes out at all 
the other points.
In this case, \emph{no} Lorentz--Lorenz local field exists
for any atom at $n > 1$, and the only field and atomic excitation
passed along the array of atoms will be locsitons.
This may be the best way to excite and observe
``pure'' locsitons, with them not being
masked by any external, averaged, mean, etc., fields.
With such an arrangement, the 1D array (or a sufficiently
thin atomic ``cylinder'' or carbon nanotube) becomes
a true and effective waveguide for locsitons,
capable of transmitting non-diffracting
radiation and atomic excitation from one location
(e.\,g.\ in opto-electronic circuits) to another.

The main issue here is how far the locsiton can propagate.
One can investigate it by studying the dispersion
relation (\ref{eq4.1}) or (\ref{eq4.2}), which predicts not 
only the locsiton wavenumbers, $q' \equiv \mathrm{Re} (q)$, 
for any given frequency detuning $\delta$,
but also their dissipation depth or distance
(in terms of numbers of atoms)
for each wavenumber, as $N_\mathrm{dis} = 1/q''$
where $q'' \equiv \mathrm{Im} (q)$.
Using only the NNA dispersion relation (\ref{eq4.2}),
amply sufficient here, since the calculation
of the dissipation distance doesn't require the same
precision as for the real wavenumbers,
we find out that the exact solutions for $q'$
and $q''$ are determined by:
\begin{eqnarray}
\cos^2 (q') &=& \frac{1}{2} 
    \left( 1 + \frac{\delta^2 + 1}{\delta_\mathrm{LL}^2} \right)
\nonumber
\\
&&{}
- \sqrt{\frac{1}{4} 
        \left( 1 + \frac{\delta^2 + 1}{\delta_\mathrm{LL}^2} \right)^2 
        - \frac{1}{\delta_{LL}^2}}
\label{eq8.1}
\end{eqnarray}
and
\begin{eqnarray}
\sinh^2 (q'') &=& -\frac{1}{2} 
    \left( 1 - \frac{\delta^2 + 1}{\delta_\mathrm{LL}^2} \right)
\nonumber
\\
&&{}
+ \sqrt{\frac{1}{4} 
        \left( 1 - \frac{\delta^2 + 1}{\delta_\mathrm{LL}^2} \right)^2 
        + \frac{1}{\delta_\mathrm{LL}^2} }.
\label{eq8.2}
\end{eqnarray}
From (\ref{eq8.1}), for very small dissipation,
i.\,e.\ when $\delta_\mathrm{LL}^2 > \delta^2 \gg 1$,
we have, as expected,
\begin{equation}
\cos (q') \approx \frac{\delta}{\delta_\mathrm{LL}}.
\label{eq8.3}
\end{equation}
The dissipation length in terms of
the numbers of atoms, $N_\mathrm{dis} = 1/q''$,
is readily found from (\ref{eq8.2}).
At the exact atomic resonance, $\delta  = 0$,
if $\delta_\mathrm{LL}^2 \gg 1$, we have:
\begin{equation}
q'' \approx \frac{1}{|\delta_\mathrm{LL}|},
\qquad
N_\mathrm{dis} \approx |\delta_\mathrm{LL}|.
\label{eq8.4}
\end{equation}
In general, in the most part of the locsiton band
(except for the edge areas, $\delta_\mathrm{LL}^2 - \delta^2 \lesssim 1$),
we find that
\begin{equation}
q'' \approx \frac{1}{\sqrt{\delta_\mathrm{LL}^2 - \delta^2}},
\qquad
N_\mathrm{dis} \approx \sqrt{\delta_\mathrm{LL}^2 - \delta^2}.
\label{eq8.5}
\end{equation}
Finally, at the locsiton band edges, defined as
$\delta_\mathrm{LL}^2 - \delta^2 = 1$,
(\ref{eq8.2}) results in
\begin{equation}
q'' \approx \frac{1}{\sqrt{\delta_\mathrm{LL}}},
\qquad       
N_\mathrm{dis} \approx \sqrt{\delta_\mathrm{LL}},
\label{eq8.6}
\end{equation}
and it remains roughly the same as $\delta^2$
reaches $\delta_\mathrm{LL}^2$.

Thus, in the most of the locsiton band,
the dissipation length, in terms of the numbers of atoms,
is $N_\mathrm{dis} = O(\delta_\mathrm{LL})$,
which also determines the maximum size  of array
to still enable the size-related resonances, 
in agreement with Section~\ref{sec5}.

Let us address now the characteristic velocities of the locsitons.
Neglecting decay in (\ref{eq4.2}) 
($\delta_\mathrm{LL}^2 > \delta^2 \gg 1$),
the group velocity of locsitons, 
$v_\mathrm{gr} = l_a (d\omega / dq)$, is found as:
\begin{equation}
v_\mathrm{gr} = \frac{l_a}{T} \frac{d\delta}{dq},
\qquad
\cos q = \frac{\delta}{\delta_\mathrm{LL}},
\label{eq8.7}
\end{equation}
hence
\begin{eqnarray}
\frac{dq}{d\delta} &=& -\frac{1}{\sqrt{\delta_\mathrm{LL}^2 - \delta^2}},
\nonumber
\\
v_\mathrm{gr} &=& l_a \sqrt{\Omega_\mathrm{R}^2 - \Delta \omega^2}
= v_\mathrm{R} \sqrt{\delta_\mathrm{LL}^2 - \delta^2},
\label{eq8.8}
\end{eqnarray}
where $\Omega_\mathrm{R}$ is the Rabi frequency of the 
self-interacting array (\ref{eq4.6}), and 
$v_\mathrm{R} = \Omega_\mathrm{R} l_a$
is a characteristic Rabi speed,
\begin{equation}
\frac{v_\mathrm{R}}{c} 
= \alpha \left| \frac{d_a}{e l_a} \right|^2 O(1) \ll 1,
\label{eq8.9}
\end{equation}
which could be even slower than a typical 
speed of sound in condensed matter.
This effect can be used e.\,g.\ for
developing nano-size delay lines in molecular computers
and in optical gyroscopes.
The LW-locsiton phase velocity is
\begin{equation}
v_\mathrm{ph} \approx \frac{v_\mathrm{R}^2}{v_\mathrm{gr}}.
\label{eq8.10}
\end{equation}
%

\section{Nonlinear excitation of a 1D array; optical bistability and hysteresis}
\label{sec9}

So far we studied linear (in field) excitations
of atomic 1D arrays.
The nonlinear interactions open door
to a huge new landscape of effects.
The simplest and very generic
nonlinearity in a two-level system is the saturation
of its absorption (\ref{eq2.4}), (\ref{eq2.5}),
which translates into a nonlinear
response of the atom polarization to the local field (\ref{eq2.6});
this, in turn, nonlinearly affects the strength of interaction
between atoms (\ref{eq2.9}).
This represents a rare case whereby
the nonlinear change (decrease) of absorption directly affects
the eigen-frequencies of the system (1D array),
by directly reducing the interaction.
Many nonlinear effects are brought up in 
short pulse modes, e.\,g.\ discrete solitons,
to be considered by us elsewhere.
However, spectacular effects, such as hysteresis
and optical bistability, emerge even in \emph{cw} mode.

To write a set of nonlinear equations for an infinite 1D array,
we first scale all fields to the 
characteristic saturation field, $E_\mathrm{sat}$ (\ref{eq2.5}),
instead of scaling to the incident field, $E_\mathrm{in}$,
so that the dimensionless local fields at the $n$-th
atom, $Y_n$, and the incident field, $X$, are
\begin{equation}
X = E_\mathrm{in} / E_\mathrm{sat},
\qquad
Y_n = E_n / E_\mathrm{sat},
\label{eq9.1}
\end{equation}
and the nonlinear counterpart of (\ref{eq3.1}) and (\ref{eq3.2})
for the array is written now as:
\begin{equation}
Y_n - \delta_\mathrm{LL}(\delta - i) 
\sum_\mathrm{latt}^{j \ne n} 
    \frac{Y_j / 2S}{|j - n|^3 (1 + \delta^2 + |Y_j|^2)} = X,
\label{eq9.2}
\end{equation}
where $\delta_\mathrm{LL}$ is defined by (\ref{eq3.4}),
so it covers either incident polarization.
The nonlinear counterpart of the NNA equation (\ref{eq3.3}) is now
\begin{eqnarray}
&&
Y_n - \frac{\delta_\mathrm{LL}(\delta - i)}{2}
\left( \frac{Y_{n-1}}{1 + \delta^2 + |Y_{n-1}|^2} \right.
\nonumber
\\
&&\qquad\qquad\left.{}
       + \frac{Y_{n+1}}{1 + \delta^2 + |Y_{n+1}|^2} \right) = X,
\label{eq9.3}
\\[3\jot]
&&
\mathcal{E}_0 = \mathcal{E}_{N+1} = 0.
\nonumber
\end{eqnarray}
None of this is an easy object even for numerical solution,
let alone analytical one.
We have, however, derived in \cite{cit1} a closed analytical solution for 
the nonlinear mode in the most fundamental system -- 
an array of just two atoms --
and found bistability and hysteresises in such a mode.
In this paper, more as a matter of illustration, 
we find an analytical multistable solution and hysteresis
for the simplest case of Lorentz--Lorenz uniform field;
however, our computer simulations showed that 
multistability and hysteresises exist in the vicinity of each
size-related resonance in the system.

For the sake of simplicity,
we consider here the case of Lorentz--Lorenz uniform mode.
In the near-neighbor approximation (\ref{eq9.3})
[the FIA results will essentially differ only by a factor of $O(1)$],
we have $Y_n = Y_{n-1} = Y_{n+1} \equiv Y$,
and thus the nonlinear equation for the uniform
local field, $Y$, is
\begin{equation}
Y \left[ 1 - \frac{\delta_\mathrm{LL}(\delta - i)}
                  {1 + \delta^2 + |Y|^2} \right] = X,
\label{eq9.4}
\end{equation}
or for the field intensity, $|Y|^2 \equiv y$,
\begin{equation}
y\, \frac{[1 - \delta (\delta_\mathrm{LL} - \delta) + y]^2 
        + \delta_\mathrm{LL}^2}{(1 + \delta^2 + y )^2} = X^2.
\label{eq9.5}
\end{equation}
It can be readily seen that the strongest nonlinear effect
emerges near the Lorentz--Lorenz resonance, 
$\delta \approx \delta_\mathrm{LL}$.
Assuming small losses, $\delta, \delta_\mathrm{LL} \gg 1$,
we can analyze the threshold of the multistability and
hysteresis mode by stipulating that $y \ll \delta_\mathrm{LL}^2$, \
$\Delta \equiv \delta_\mathrm{LL} - \delta \ll \delta_\mathrm{LL}$,
and thus (\ref{eq9.5}) can be further simplified as
\begin{equation}
y [(y - \delta_\mathrm{LL} \Delta)^2 + \delta_\mathrm{LL}^2] 
    / \delta_\mathrm{LL}^4 \approx X^2.
\label{eq9.6}
\end{equation}
The threshold for the
multistable solution $y(X)$ of this equation 
is determined by the condition $dX / dy = d^2 X / d y^2 = 0$,
which results in the critical requirement that
\begin{eqnarray}
\Delta \equiv \delta_\mathrm{LL} - \delta 
&>& \Delta_\mathrm{cr} = \sqrt 3,
\nonumber
\\    
y \equiv |Y|^2 &>& y_\mathrm{cr} 
= \frac{2}{\sqrt{3}} \delta_\mathrm{LL},
\label{eq9.7}
\\
X^2 &>& X_\mathrm{cr}^2 = \frac{1}{\delta_\mathrm{LL}^2}
    \left( \frac{2}{\sqrt{3}} \right)^3.
\nonumber
\end{eqnarray}
Amazingly, as one recalls that $X = E_\mathrm{in} / E_\mathrm{sat}$ and
$\delta_\mathrm{LL} \gg 1$, the critical (threshold) driving
intensity $E_\mathrm{in}^2$ to initiate 
multistability and hysteresis could be orders of 
magnitude lower than the saturation one, $E_\mathrm{sat}^2$,
which is mostly due to resonant nature of the effect
that emerge in the vicinity of the Lorentz--Lorenz resonance.
Thus, the required saturation nonlinearity is indeed
just a slight perturbation to the resonant linear mode.
This should be of no big surprise,
since the nature of the effect is the same as in 
many other so called 
\emph{vibrational hysteresises} \cite{cit13}
in resonant nonlinear systems, from pendulum \cite{cit13}
to electronic circuits \cite{cit14},
to optical bistability in a Fabri--Perot resonator \cite{cit15},
to a cyclotron resonance of a slightly relativistic 
electron \cite{cit16}.
In all these nonlinear resonances, it is enough for
a narrow resonant curve to be tilted 
beyond its resonance width to reach
multistability by becoming a ``Pisa-like tower''.

\begin{figure}
\includegraphics[width=8.6cm]{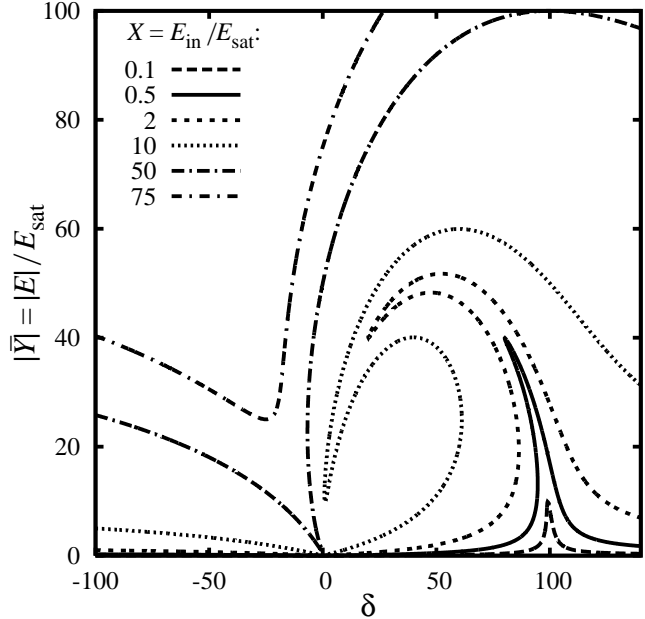}
\caption{%
Multistability of Lorentz--Lorenz mode in a long
array: local Lorentz field amplitude $Y$ \emph{vs} detuning
$\delta$ for different normalized driving
amplitudes $X = E_\mathrm{in} / E_\mathrm{sat}$
for the case of $\delta_\mathrm{LL} = 100$.}
\label{fig5}
\end{figure}

The formation of bistability and hysteresis
is depicted in Fig.~\ref{fig5}, where the resonant curve
$Y(\delta)$ is shown for different driving $X$
for the full equation (\ref{eq9.5}).
The formation of a tri-stable solution
results in the so called bistability, whereby
only the states with the maximum and minimum
intensities are stable, whereas the middle
branch of the solution is absolutely unstable.
To an extent, it is similar to one of the hysteretic patterns 
for the local field found for the case of 
the local field \cite{cit1} and scattered light \cite{cit17new}
of just two atoms; the hysteretic resonance
here is produced by a Lorentz-like, or ferromagneticlike, 
excitation of atoms with LW locsitons.
This effect is also reminiscent of optical intrinsic bistability
for uniform LF in dense materials
predicted earlier in \cite{cit17} and observed experimentally 
in \cite{cit18}.

The above result for hysteresis and bistability 
was found for the uniform, Lorentz, field,
i.\,e.\ near the Lorentz--Lorenz resonance,
using a simple field distribution.
However, it is clear from the above
argument with a ``Pisa-like tower'', a
similar hysteretic effect can be expected 
in the close vicinity of \emph{every} size-related
locsiton resonance within the locsiton band.
Some of the most interesting hysteresises, though,
are excited for the SW locsitons, when
the nearest atoms counter-oscillate
in an antiferromagneticlike fashion, as shown in \cite{cit1}
for the anti-Lorentz band edge for two atoms.
This hysteresis results in a ``split-fork'' bistability
and will be discussed by us in application to long arrays elsewhere.

\section{Analogies to locsitons in other physical systems and general 
discussion}
\label{sec10}

It is only natural to expect the effects studied here
to manifest themselves in other finite systems of 
discrete resonant oscillators that are externally driven and 
interact with each other.
On the one hand, this may help 
to demonstrate and verify some major results
reported here in much simpler 
and easily handleable settings;
on the other hand, this may bring up new features
in these other systems that escaped attention in 
seemingly well developed and researched fields.
We consider here a few such systems.

Perhaps, the most classical example would be a mechanical
finite array of identical physical pendulums with the
same individual resonant frequencies, weakly coupled
to each other (say, by using weak strings 
between neighboring pendulums), with each pendulum driven
independently by an external feed (say, \emph{via} EM solenoids)
with the same phase for all of them.
By tuning the frequency of the driving feed around
the resonant frequency of the pendulums, one may expect
to observe stratified excitation of the pendulums,
from long-wave, ferromagneticlike strata, to
short-wave, antiferromagneticlike ones.
One can also expect to observe ``magic'' 
numbers and related effect of non-moving pendulums
surrounded by strongly excited ones.
(It is worth noting that we are talking here about a \emph{cw}
motion, in contrast to the well known effect
in two coupled non-driven pendulums, whereby
the pendulums periodically alternate periods 
of zero-excitation in one and strong excitation in the other one,
with the frequency of the alternation being inversely
proportional to the coupling.)
The major effect here would be again due 
to locsitons, and the major critical condition 
for their excitation would be similar
to the condition on the strength of interaction,
$Q_a$, (\ref{eq2.10}), to exceed a critical one.

The driven pendulum array would be a great classroom-demonstration tool. 
However, although purely mechanical, it might take some effort to
implement, due to the need of an independent feed to each pendulum.
From that point of view, a much simpler, easily 
implementable, controllable, recordable,
and versatile system could be an electronic
array of individual resonant circuits,
electronically coupled to each other.
A long while ago, similar systems have
been used as transmission lines, whereas
a finite set of circuits has also been of 
interest for such applications as bandpass filters.
However, the detailed picture of behavior of
individual circuits inside such systems
have apparently not attracted too much attention,
with a few reasons for that.
The main difference between these
systems and the ones proposed by us, is that
we need an independent feed with the same phase
for each individual circuit, 
which may be arranged, by e.\,g.\ using 
individual cable for each one of them.
The coupling between the individual circuits
can be engineered in such a way that
one can arrange strictly near-neighbor
interaction, two-near-neighbors interaction, etc.
Besides, the contribution of each neighboring
circuit can be independently controlled,
so the term $1/n^3$ can now be changed
to any desirable function of the neighbor spacing.

A close example, which might have important
implications for large radio-frequency  antenna arrays,
for e.\,g.\ radio-astronomy applications,
as well as in multi-dish radar systems,
is the interaction of radiators in those arrays,
if the strength of this interaction exceeds some critical value.

Since in the case of electronic circuits the polarization
of the feed and the spatial anisotropy as in (\ref{eq2.2})
is not a factor anymore, the dipole-like interaction
is simplified, as its spatial polarization form-factor $F_{(p)}$
(as in Section~\ref{sec3}) can be now dropped,
and the equations of motion (\ref{eq3.1})--(\ref{eq3.4}) 
can be simplified.
Furthermore, since now the direction of a ``dipole''
with respect to the ``incident polarization''
is not a factor, one can arrange a \emph{loop} or \emph{ring}
of those circuits, instead of a linear 1D array,
which allows for periodic boundary condition
instead of zero boundary conditions as e.\,g.\ in (\ref{eq3.1}).
This would greatly simplify the theory and comparison
with the experiment, on the one hand,
and allow for new interesting effects in ring arrays
(such as e.\,g.\ greatly enhanced Sagnac effect and
related gyroscope applications),
to be discussed by us elsewhere, on the other hand.

An interesting and exotic opportunity with 1D arrays, 
and especially ring arrays, is a possibility
to develop a toy model of ``discrete space quantum mechanics'',
whereby a wave function is replaced by a set of 
oscillating fields at discrete spatial points, 
instead of a regular wave distributed in space.
This discrete-space quantum mechanics (QM) could be of interest
for the theory of certain systems,
as well as from fundamental point of view;
for example, the ring arrays then would allow to 
build a theory of a Bohr-like ``discrete-QM H atom'',
with finite number of primary quantum numbers,
unlike an infinite numbers of quantum levels 
as in a regular-QM H atom.

A very interesting analogy, 
especially from the application point of view,
is \emph{magnetic spin resonance}, in particular
\emph{nuclear magnetic resonance} (NMR) \cite{cit19},
as well as magnetic resonances in finite spin systems, in particular 
so called molecular magnets \cite{cit20}.
They have some common points with the interacting systems 
considered here, in particular, two-level nature of resonances
in both systems.

In optical domain, an observation of the effect discussed 
here can be done in a few ways.
Nanostrata and locsitons can be observed 
either \emph{via} size-related resonances
in scattering of laser radiation,
or \emph{via} X-ray or electron 
energy loss spectroscopy of the strata.

The effect has promising potential for molecular computers
\cite{cit21} and nanodevices.
The major advantage of locsitons \emph{vs} electrons in semiconductors
is that they are not based on electric current or charge transfer.
This may allow for a drastic reduction of the size limit for
computer logic elements currently based on 
metal-oxide-semiconductor technology, which may
suffer from many irreparable problems on a scale below 10~nm.
As such, locsiton-based devices
could be an interesting entry into the field, as
complimentary or alternative to emerging technologies
like plasmonics \cite{cit22} or spintronics \cite{cit23}.
They can offer both passive
(e.\,g.\ transmission lines and delays),
and active elements, e.\,g.\ for switching and logics.
A ring-array may be used as a basis
for a Sagnac-locsiton-based gyroscope;
low locsiton velocity
may allow for a high sensitivity in a small ring.

Another promising application of locsitons could be
biosensing devices, where
target-specific receptor molecules either form
a locsiton-supporting lattice
or are attached to its sites;
a localized locsiton occurs
whenever a target biomolecule attaches to a receptor.

Finally, exciting opportunities
exist in atomic arrays and lattices with inverse
population created by an appropriate (e.\,g.\ optical)
pumping, which may lead to a laser-like
locsiton stimulated emitter (``locster''), 
to be discussed elsewhere.

\section{Conclusions}
\label{sec11}

In conclusion, in this study of strong 
stratification of local field 
and dipole polarization in finite groups of atoms
predicated by us earlier \cite{cit1},
we developed a detailed theory of the phenomenon in one-dimensional
arrays of atom or resonant particles.
In strong departure from  Lorentz--Lorenz theory,
the spatial period of those strata 
may become much shorter than the incident wavelength.
By exploring  nanoscale elementary excitations, locsitons,
and resulting size-related resonances and large field 
enhancement in finite arrays of atoms, we showed that
their spatial spectrum has both
long waves, reminiscent of ferromagnetic 
domains, and super-short waves corresponding to
the counter-oscillating neighboring 
polarizations, reminiscent of antiferromagnetic spins.
The system also exhibits ``hybrid'' modes of excitation
that have no counterpart in magnetic ordering,
and are more representative of the effect.
Our theory that goes beyond
Ising-like near-neighbor approximation and describes
the excitation whereby each atom interacts with all 
the other atoms in the array,
reveals the existence of infinite spectrum of ``exponential'', or  
``evanescent'' eigenmodes in the such arrays.
We explored the phenomenon of ``magic'' numbers of atoms 
in an array, whereby resonant local-field suppression
can be canceled for certain atoms in an array.
We also demonstrated the existence of nonlinearly
induced optical bistability and hysteresis in the system.
We discussed a stratification effect
similar to that in atomic arrays, which may exist in 
broad variety of self-interacting
systems, from mechanical (pendulums)
to electronic circuits, to radar arrays,
and to the nuclear magnetic resonance.
We pointed out a few potential applications
of the atomic and similar arrays in such diverse field
as low-losses nano-elements for optical
computers, small-size Sagnac-effect-based gyroscopes,
and bio-censors.

\section{Acknowledgements}

The authors appreciate helpful discussions with V.~Atsarkin
and R.~W.~Boyd.
This work is supported by AFOSR.

\appendix
\section{Mathematical aspects of the dispersion equation (\ref{eq4.1})}
\label{app}

In this Appendix we consider mathematical
properties of the Fourier series in the dispersion 
relation (\ref{eq4.1}), i.\,e.\ the function 
\begin{equation}
\Sigma_\rho (q) 
\equiv \sum_{n=1}^\infty \frac{\cos(nq)}{n^\rho},
\label{eqA.1}
\end{equation}
which is a generalized form of (\ref{eq4.1}) [in (\ref{eq4.1}),
$\rho = 3$ due to the chosen geometry of the problem --
1D array of point dipoles],
but we will restrict ourselves to natural numbers $\rho$.
Note, for example, that in the case of infinite ``dipole strings'',
parallel to each other and equidistantly arranged in
a 2D plane, $\rho = 2$, whereas for
``dipole planes'' parallel to each other and equidistantly arranged in
the 3D space, $\rho = 1$.

Our main purpose here is to find a closed finite analytical
form of the function $\Sigma_\rho (q)$ that originated the 
Fourier series (\ref{eqA.1}), and when it is impossible
in terms of more or less regular analytical
functions, at least find a closed finite
derivative $d^m \Sigma_\rho (q) / d q^m$
with minimum derivative order $m$,
which will help us to analyze in great detail
the behavior of the function $\Sigma_\rho (q)$
in its physically interesting points,
e.\,g.\ at $q = 0, \pi$, the ratio
of its min/max values 
$\Sigma_\rho (\pi) / \Sigma_\rho (0)$, etc.
Those derivatives are actually the main
tool in our search.
Indeed, differentiating $\Sigma_\rho (q)$ in (\ref{eqA.1})
$\rho - 1$ times,
one obtains
\begin{eqnarray}
\frac{d^{\rho-1} \Sigma_\rho (q)}{d q^{\rho-1}} 
&=& (-1)^{\rho/2} \sum_{n=1}^\infty \frac{\sin(nq)}{n},
\label{eqA.2}
\\
&&\qquad\qquad\mbox{if}\quad\rho = 2, 4, 6, \ldots
\nonumber
\end{eqnarray}
and
\begin{eqnarray}
\frac{d^{\rho-1} \Sigma_\rho (q)}{d q^{\rho-1}} 
&=& (-1)^{(\rho-1)/2} \sum_{n=1}^\infty \frac{\cos(nq)}{n}
\label{eqA.3}
\\
&=& (-1)^{(\rho-1)/2} \Sigma_1 (k),
    \qquad\mbox{if}\quad  \rho = 1, 3, 5, \ldots
\nonumber
\end{eqnarray}
Recalling that $\sin \xi = (e^{i\xi} - \mathrm{c.c.})/2i$
and $\cos \xi = (e^{i\xi} + \mathrm{c.c.})/2$,
we recognize the sums
$\sum_{n=1}^\infty e^{\pm inq} / n$
in (\ref{eqA.2}) and (\ref{eqA.3}) 
as Taylor expansions of $-\ln(1 - e^{\pm iq})$.

In the case of even numbers $\rho$, the sum (\ref{eqA.2})
in the interval $0, 2\pi$ (which is of the most interest) yields:
\begin{equation}
\sum_{n=1}^\infty \frac{\sin(nq)}{n} = 
\begin{cases}
\displaystyle
\frac{\pi \mathop{\mathrm{sign}}(q) - q}{2},
&\quad\mbox{if}\quad  |q| \in (0, 2\pi),
\\
0, &\quad\mbox{if}\quad  |q| = 0, 2\pi.
\end{cases}
\label{eqA.4}
\end{equation}
It is discontinuous at $q = 0, \pm 2\pi$.
Even numbers $\rho$ are thus a ``lucky'' case;
the result (\ref{eqA.4}) is easily integrated to restore
function $\Sigma_\rho (q)$. In a physically interesting case,
$\rho = 2$, by integrating (\ref{eqA.4}) once we have
\begin{eqnarray}
\Sigma_2 (q) &\equiv& \sum_{n=1}^\infty \frac{\cos(nq)}{n^2}
= \frac{3(\pi - |q|)^2 - \pi^2}{12},
\label{eqA.5}
\\
&&\qquad\qquad\mbox{if}\quad  q \in [-2\pi, 2\pi],
\nonumber
\end{eqnarray}
which is a continuous (but not smooth) function with
$\Sigma_2 (0) = \pi^2/6$ and 
$\Sigma_2 (\pi) = -\pi^2/12$,
so that $(\Sigma_2)_\mathrm{min} / (\Sigma_2)_\mathrm{max} = -1/2$.

As to the closed integrability, one is not as lucky with
odd numbers $\rho$; 
the summation of (\ref{eqA.3}) yields:
\begin{eqnarray}
&&
\Sigma_1 (q) \equiv \sum_{n=1}^\infty \frac{\cos(nq)}{n}
= -\ln[2\,|\sin(q/2)|],
\label{eqA.6}
\\
&&\quad\mbox{i.\,e.}\quad  \Sigma_1 (q) \to \infty
\quad\mbox{at}\quad  q = 2m\pi,
\nonumber
\end{eqnarray}
which cannot be integrated in simple known analytical
functions, but it gives a good analytical tool to analyze
the behavior of $\Sigma_\rho (q)$.
In the case of most interest to us, 
$\rho = 3$, we have that near the maximum
of $\Sigma_3$ at small wavenumbers
\begin{equation}
\Sigma_3(q) \approx S + \frac{q^2}{2} \left( \ln|q| - \frac{3}{2} \right)
\qquad\mbox{at}\quad  |q| \ll 1,
\label{eqA.7}
\end{equation}
where $S = \sum_{j=1}^\infty j^{-3} \approx 1.202057$,
whereas near the minimum of $\Sigma_3$, i.\,e.\
near $q = \pi$, we have
\begin{equation}
\Sigma_3(q) \approx -\frac{3S}{4} + \frac{(q - \pi)^2}{2} \ln 2,
\qquad\mbox{if}\quad  |q - \pi| \ll 1.
\label{eqA.8}
\end{equation}
Notice that here 
$(\Sigma_3)_\mathrm{min} / (\Sigma_3)_\mathrm{max} = -3/4$,
and $d^2 \Sigma_3 / d q^2 = 0$ at
$q = 2m\pi \pm \pi/3$.

For the magic numbers in the case of full interaction
of individual atoms with all the other atoms in the array,
it is important to know zeros of the function $\Sigma_3(q)$.
As it has been mentioned before [see (\ref{eq7.2}) and related text],
it turns out that the value of the ratio $q'/\pi$ 
for the first positive root of the equation $\Sigma_3(q) = 0$
is very close to a small rational number:
\begin{equation}
\frac{q'}{\pi} = (1 + \Delta') \cdot \frac{6}{13},
\qquad\mbox{with}\quad  \Delta' \approx 2.6 \times 10^{-4}.
\label{eqA.9}
\end{equation}
For all practical purposes, a good approximation
for $\Sigma_3(q)$ is provided by:
\begin{eqnarray}
&&
\Sigma_3^\mathrm{(fit)} (q) 
= S \left\{ \cos q 
\phantom{\frac{1}{1}}\right.
\nonumber
\\
&&\quad\left.{}
+ \frac{\ln[(S + \Delta S) |\sin(q/2)|]}
                          {\ln(S + \Delta S)} 
            \cdot \frac{\sin^2 (q/2)}{4} \right\},
\label{eqA.10}
\end{eqnarray}
where $\Delta S = 0.01472$.
It coincides with the results provided by numerical
summation of (\ref{eqA.1}) with $\rho = 3$ with the precision
better than $0.6\%$ of $S$, and their zeros coincide
with each other and with (\ref{eqA.9}) with precision 
better than $10^{-6}$.

Let us also prove the relation for
$(\Sigma_{\rho})_\mathrm{min} / (\Sigma_{\rho})_\mathrm{max}$ 
used in (\ref{eq4.4}).
Indeed,
\begin{eqnarray}
&&
\frac{D(\pi,\rho)}{D(0,\rho)} 
= \frac{(\Sigma_{\rho})_\mathrm{min}}{(\Sigma_{\rho})_\mathrm{max}}
= \frac{\displaystyle \sum_{n=1}^\infty (-1)^n n^{-\rho}}
       {\displaystyle \sum_{n=1}^\infty n^{-\rho}} 
\nonumber
\\
&&{}=
\frac{\displaystyle 
      -\sum_{n=1}^\infty n^{-\rho} + 2 \sum_{m=1}^\infty (2m)^{-\rho}}
     {\displaystyle 
      \sum_{n=1}^\infty n^{-\rho}} 
\label{eqA.11}
\\
&&{}=
\frac{\displaystyle 
      -\sum_{n=1}^\infty n^{-\rho} + 2^{-\rho+1} \sum_{n=1}^\infty n^{-\rho}}
     {\displaystyle 
      \sum_{n=1}^\infty n^{-\rho}} 
= -1 + \frac{1}{2^{\rho-1}}.
\nonumber
\end{eqnarray}
Note that (\ref{eqA.11}) is valid for \emph{any} number $\rho > 1$.


\end{document}